\documentclass[journal]{IEEEtran}
\ifCLASSINFOpdf
\else
\fi
\usepackage{array}

\usepackage{algorithm}
\usepackage{algpseudocode}

\usepackage{cite}
\usepackage{amsmath,amssymb,amsfonts}
\usepackage{graphicx}
\usepackage{textcomp}

\usepackage[utf8]{inputenc}
\usepackage{physics}
\usepackage{lineno}
\usepackage{cite}
\usepackage{enumerate,subfigure}
\usepackage{latexsym}
\usepackage{framed}
\usepackage{hyperref}
\usepackage{cite}
\usepackage{afterpage}
\usepackage{comment}
\usepackage{multirow}

\usepackage[dvipsnames]{xcolor}

\DeclareMathOperator{\R}{{\mathbb R}}

\DeclareMathOperator{\texp}{\mathbf{t}^{\mbox{\tiny{\textbf{exp}}}}}

\DeclareMathOperator{\xrec}{x_{\mbox\tiny recon}}
\DeclareMathOperator{\xreci}{x^i_{\mbox\tiny recon}}
\DeclareMathOperator{\xin}{x_{\mbox\tiny in}}

\DeclareMathOperator{\xtruei}{x^i_{\mbox\tiny true}}

\begin{document}

%
\title{Domain independent post-processing with graph U-nets:  Applications to Electrical Impedance Tomographic Imaging }
%
%
%

\author{William Herzberg, Andreas Hauptmann~\IEEEmembership{Senior Member,~IEEE}, and Sarah J. Hamilton
\thanks{SH and BH were supported by the National Institute Of Biomedical Imaging And Bioengineering of the National Institutes of Health under Award Number R21EB028064. The content is solely the responsibility of the authors and does not necessarily represent the official views of the National Institutes of Health.  AH was supported by Academy of Finland projects 338408, 346574, 353093.
SH and AH would like to thank the Isaac Newton Institute for Mathematical Sciences, Cambridge, for support and hospitality during the programme \emph{Rich and Nonlinear Tomography} where work on this paper was undertaken, supported by EPSRC grant no EP/R014604/1. }
\thanks{WH was with the Department of Mathematical and Statistical Sciences; Marquette University, Milwaukee, WI 53233 USA}
\thanks{AH is with the Research Unit of Mathematical Sciences, University of Oulu, Finland and also with the Department of Computer Science, University College London, UK.}
\thanks{SJH is with the Department of Mathematical and Statistical Sciences; Marquette University, Milwaukee, WI 53233 USA, \texttt{(sarah.hamilton@mu.edu)}}}

\maketitle

\begin{abstract}
Reconstruction of tomographic images from boundary measurements requires flexibility with respect to target domains. For instance, when the system equations are modeled by partial differential equations the reconstruction is usually done on finite element (FE) meshes, allowing for flexible geometries. Thus, any processing of the obtained reconstructions should be ideally done on the FE mesh as well. For this purpose, we extend the hugely successful U-Net architecture that is limited to rectangular pixel or voxel domains to an equivalent that works flexibly on FE meshes. To achieve this, the FE mesh is converted into a graph and we formulate a graph U-Net with a new cluster pooling and unpooling on the graph that mimics the classic neighborhood based max-pooling. We demonstrate effectiveness and flexibility of the graph U-Net for improving reconstructions from electrical impedance tomographic (EIT) measurements, a nonlinear and highly ill-posed inverse problem. The performance is evaluated for simulated data and from three measurement devices with different measurement geometries and instrumentations. We successfully show that such networks can be trained with a simple two-dimensional simulated training set and generalize to very different domains, including measurements from a three-dimensional device and subsequent 3D reconstructions.



\end{abstract}


%
\IEEEpeerreviewmaketitle

\section{Introduction}\label{sec:intro}

\IEEEPARstart{N}{onlinear} inverse problems are often described by partial differential equations (PDEs) and measurements are taken directly on the boundary of the domain, resulting in varying domain shapes \cite{mueller2012linear}. Consequently, reconstruction algorithms need to offer the flexibility to operate on these varying domains. The finite element method (FEM), and in particular the corresponding meshes, offers this flexibility with respect to domain shapes and hence the tomographic image is usually computed on a target specific mesh. A popular class of reconstruction algorithms for this task are optimization-based variational methods \cite{kaltenbacher2008iterative}, where the reconstructions are iteratively updated on the mesh by some optimization algorithm, such as a Gauss-Newton type method. Unfortunately, these methods tend to be expensive due to costly Jacobian computations resulting in a tradeoff between cost and image quality for increasing iterations. Additionally, reconstructions can be sensitive to modeling of the domains or measurement devices, potentially causing severe reconstruction artifacts \cite{hamilton2019comparing}.  

One way to improve the image quality from an early iterate is to perform a post-processing step to provide image quality comparable to the full iterative algorithm, but with a substantial reduction in computational cost. For this specific task Deep Learning approaches have been immensely popular in recent years \cite{arridge2019solving}. Here, given the suboptimal reconstruction from an early iterate one trains a neural network with representative training data to produce an improved reconstruction. For this specific purpose of post-processing, U-Net architectures \cite{Ronneberger2015, Gao2019} have been immensely successful.  U-nets use a multi-scale convolutional neural network (CNN) to process images on multiple resolutions by extracting edge information as well as long range features.  The main limitation of CNN U-nets lies in the strict geometric assumptions on the mesh, i.e., the application of the convolutional filters requires a regular rectangular mesh with ideally isotropic pixel dimensions, preventing their direct application to the aforementioned reconstruction problem on FE meshes. A simple remedy would be to interpolate between the FE meshes and the rectangular grid for application of the CNN, losing the flexibility that FEM provides \cite{mozumder2021modelbased}. 

Alternatively, to retain the flexibility one can interpret the FE mesh as a graph and process the image directly using graph convolutional neural networks \cite{Herzberg2021}.  Here we propose an extension of the CNN based U-Net architecture to a graph U-Net. New graph-based pooling operations are required to move between the multiple resolutions of a U-net such that the local relations are preserved. To achieve this we propose a cluster based pooling and unpooling that provides comparable down and up-sampling to the classic max, or average, pooling on CNNs. For computational feasibility, the clusters are pre-computed for each mesh and can be efficiently applied during training and inference. The proposed graph U-Net is then applied in the context of electrical impedance tomography (EIT) a highly nonlinear inverse problem that requires strong regularization to obtain good image quality. In this work, we compute an initial reconstruction with only a few iterations of total variation regularized Gauss-Newton method  and then train the network to improve image quality providing excellent reconstruction quality with a considerable reduction in processing time.

The primary advantage of the graph U-Net lies in the flexibility with respect to measurement domains. While each device and domain naturally requires their own careful modeling, we would ideally train the networks on general, simple, measurement setups.  This overcomes drawbacks, 1) the creation of general enough training data is a time intensive task and 2) we cannot predict all possible encountered domains in the measurement process, e.g., varying chest shapes for physiological measurements of human subjects \cite{sousa2014vivo}. Furthermore, the graph based nature of the network is dimension independent as neighboring nodes are described by a dimension independent adjacency matrix. This allows one to train the network in 2D and apply it to measurements from 3D domains. 

This paper addresses the challenging task to process EIT reconstructions from a diverse set of measurement devices with a single network trained in a single 2D chest-shaped measurement domain with elliptical inclusions. We can show that the network successfully generalizes to measurement data under varying domain shapes as well as to reconstructions from three different EIT devices (KIT4, ACT3, and ACT5).  The data from the KIT4 device \cite{Kournunen2008, Hamilton2019beltrami} was taken on a chest-shaped tank with 16 electrodes using bipolar current injection whereas the ACT3 data was taken on a circular tank with simultaneous injection/measurement across all 32 electrodes \cite{Cook1994,Isaacson2004}.  Both the KIT4 and ACT3 datasets corresponded to 2D cross-sectional imaging.  However, the ACT5 datasets \cite{rajabishishvan2021,Hamilton2022} used a fully 3D box geometry, with 32 large electrodes and simultaneous current injection. 

This work is the first to show that a single network can be successfully used in a variety of instrumentation and measurement setups, combining the flexibility of FEM  with graph neural networks (GNN) and thus making deep learning techniques more accessible to inverse problems and imaging applications that heavily rely on the use of FE meshes. For this purpose, we also provide a code package GN4IP\footnote{Graph Networks for Inverse Problems (GN4IP) is available at \url{github.com/wherzberg/GN4IP}}.  It should be noted that this paper reuses some content from thesis \cite{HerzbergThesis2022}, with permission.

The remainder of the paper is organized as follows.  Section~2 develops the graph U-net and novel pooling layers, a brief overview of EIT, the experimental data, training data, and  metrics to be used to assess reconstruction quality.  Results are given in Sec.~\ref{sec:methods} and discussion follows in Sec.~\ref{sec:discussion} where modifications and extensions are explored, including the 3D data with reconstructions from different algorithms than the network was trained on.  Conclusions are drawn in Sec.~\ref{sec:conclusions}.

\section{Methods}\label{sec:methods}

\subsection{Learned Reconstruction}\label{sec:LearnedRecon}
The main focus of this work is post-processing tasks, however the modified graph U-net and new pooling layers could easily be used in place of a residual network in a model-based learning framework such as \cite{Herzberg2021}. 
The problem at hand is to improve a fast, reliable image that has predictable artifacts that could be removed via post-processing.  Post-processing EIT reconstructions with the traditional U-net architecture was shown highly effective for d-bar based reconstructions \cite{Hamilton2018, Hamilton2019} and the dominant current scheme \cite{wei2019dominant}, but each of those works applied the networks to rectangular pixel image data, despite measurements obtained on different domains.  However, many image reconstruction methods are performed on irregular meshes (e.g. FE meshes), which then require the solution (reconstructed image) to be interpolated from the computational mesh to the pixel grid.  In some large scale cases, this may be cost-prohibitive or less desirable when solutions are needed at high precision.  Thus, it is of interest to have an alternative network structure for learned reconstruction (e.g. post-processing and model-based learning) on the computational mesh.  

\begin{figure}[h!]
	{\includegraphics[width=0.5\textwidth]{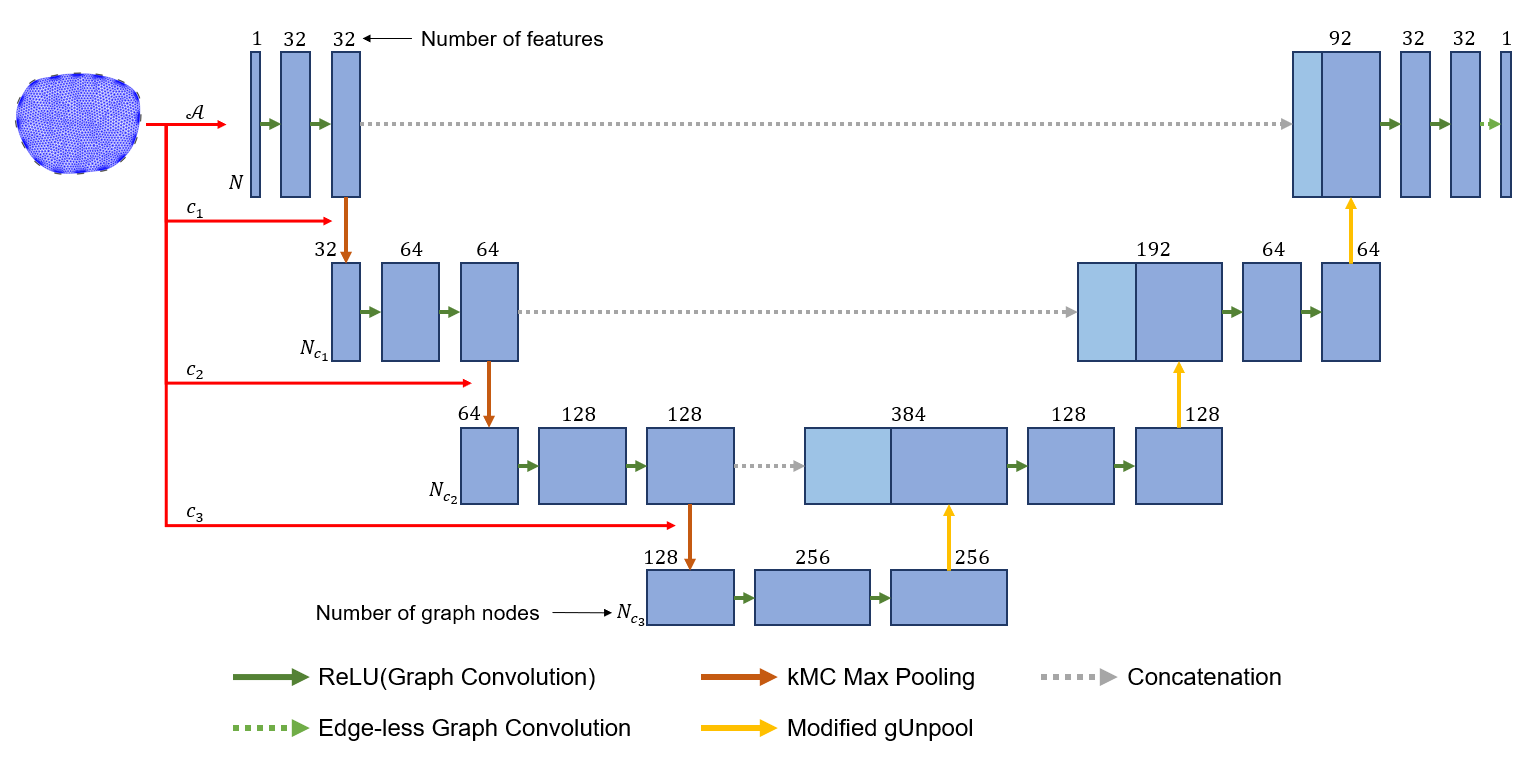}} \caption{A diagram of the proposed graph U-net with three pooling layers. The input to the network is an initial reconstruction along with the adjacency matrix $\mathcal{A}$ and cluster assignments $(c_1, c_2, c_3)$ for each of the pooling layers, determined from the mesh. The output is defined on the same graph structure as the input.}
	\label{fig:diagrams_unet_gnn}
\end{figure}

Graph Neural Networks have been around for years and recently have garnered renewed interest for their scalability and data flexibility.  As with traditional CNNs, several options for convolutional and pooling layers exist.  However, at the time of this work, the existing pooling layers, in particular were inadequate for mirroring the maxpooling in classic CNN U-nets.  Therefore, we developed new layers here based on spatially clustering neighboring nodes in the computational mesh and then performing the maxpooling over the clusters.  The structure of the modified graph U-net used here is shown in Figure~\ref{fig:diagrams_unet_gnn}.

\subsubsection{A graph U-net with cluster pooling}\label{sec:gnns}
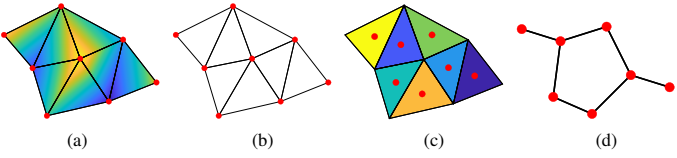
\begin{figure}[h!]

%
%
	\centering
	\begin{picture}(270,50)
	
	\put(130, 10){\includegraphics[width=60pt]{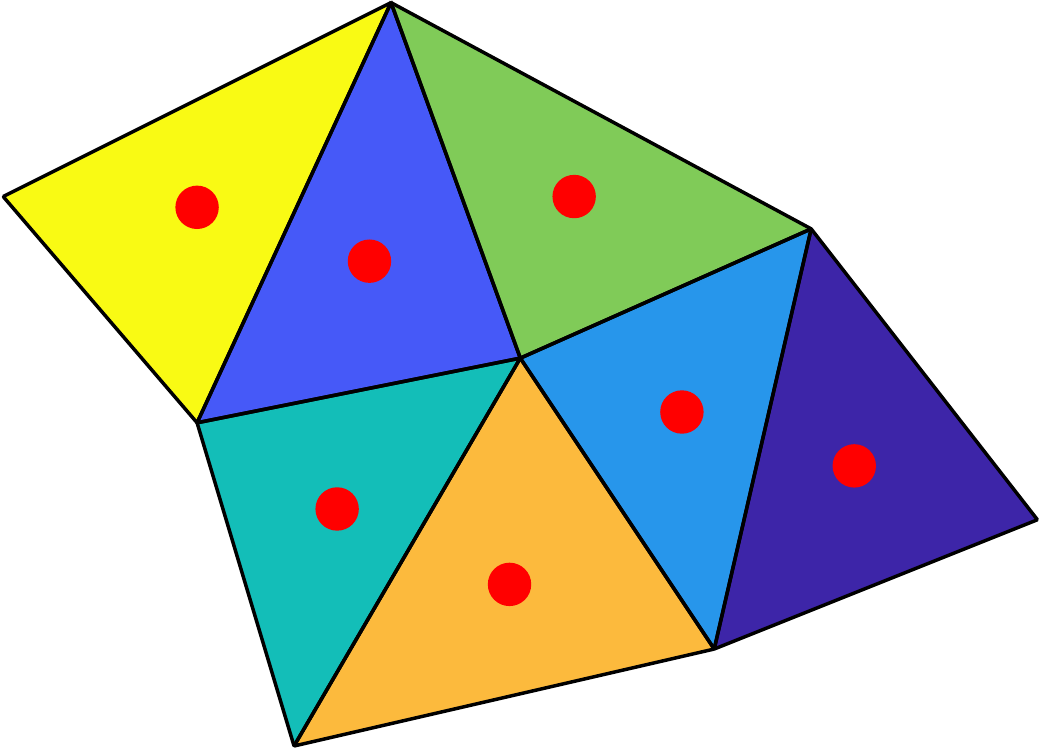}}
	\put(195, 10){\includegraphics[width= 60pt]{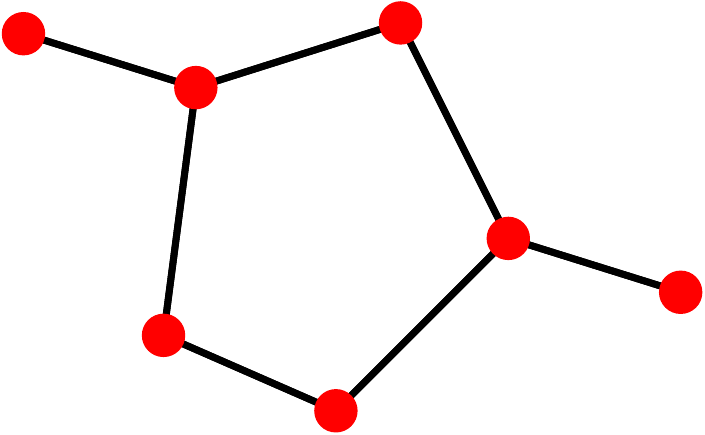}}
	
	\put(0,10){\includegraphics[width=60pt]{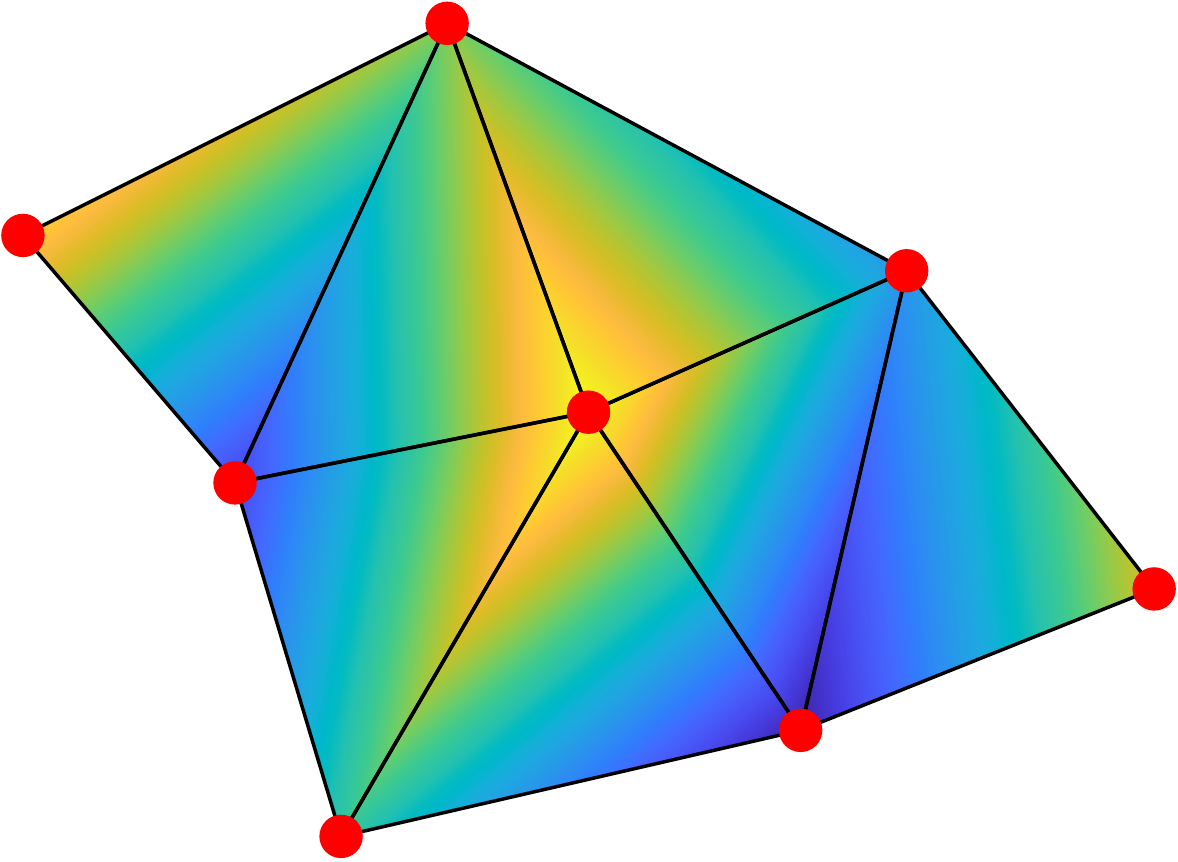}}
	\put(65,10){\includegraphics[width=60pt]{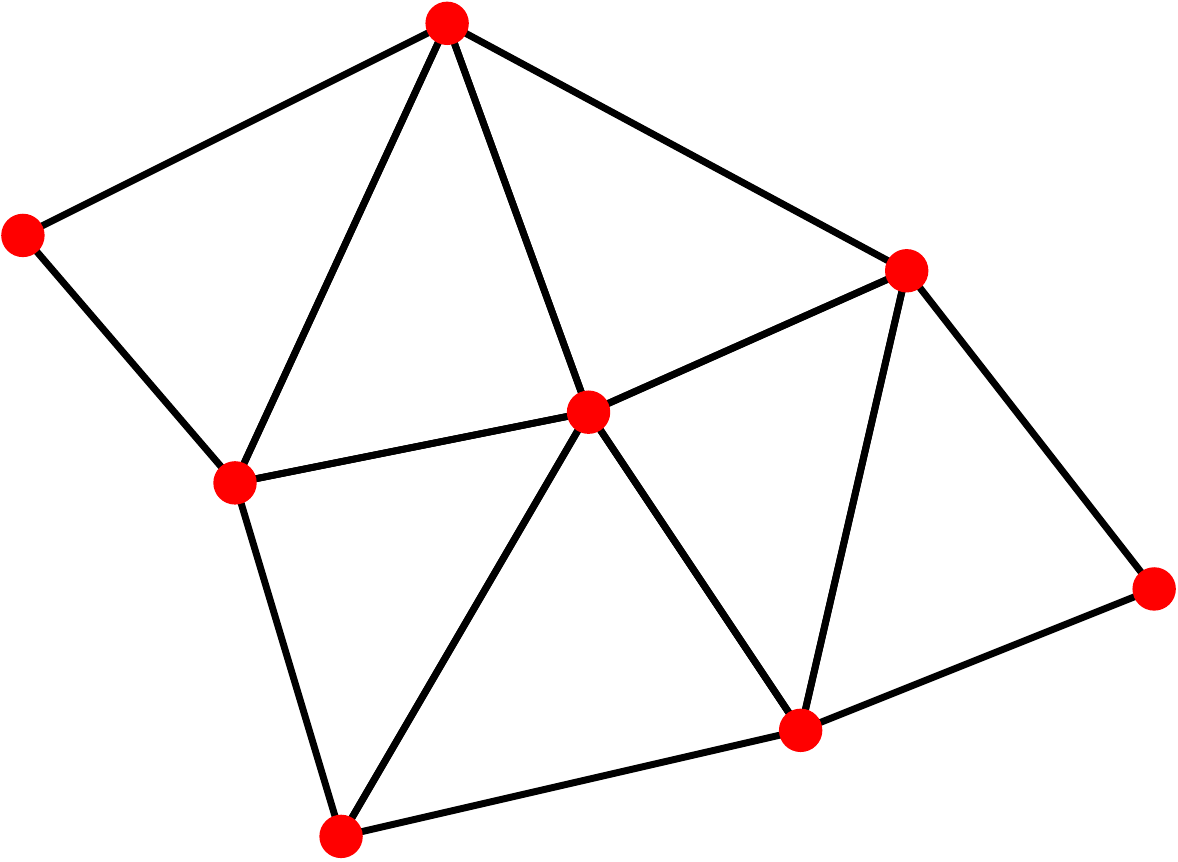}}
	
	\put( 25,0){\scriptsize(a)}
	\put(95,0){\scriptsize(b)}
	\put(160,0){\scriptsize(c)}
	\put(225,  0){\scriptsize(d)}
	
	\end{picture}
	\caption{A mesh with data defined over mesh nodes (a)  and (b) the corresponding graph.  A mesh with data defined over mesh elements (c) and (d) the corresponding graph.}\label{fig:FEMgraphs}
\end{figure}

As the name suggests, graph networks act on {\em graph} input data.  Historically, graph convolutional networks have been applied to citation networks, social networks, or knowledge graphs. However, whenever you have a computational mesh, you inherently have a graph representing the connections between the elements or nodes in your computational mesh.  In particular, for irregular meshes commonly associated with FEM there are two natural options for the associated graph: the mesh elements, or the mesh nodes (Fig.~\ref{fig:FEMgraphs}).  The {\em adjacency matrix} $\mathcal{A}$ is a sparse matrix listing the edges (connections) between the graph nodes.  The basic form of the sparse adjacency matrix is $\mathcal{A}_{ij}=1$ if there is an edge connected nodes $n_i$ and $n_j$.  Self-loops are recorded separately. 


As opposed to convolutional layers which require regular $n$-dimensional regular input data, {\em graph convolutional} layers are designed to work on simple, homogeneous graph-type data.  While the list of graph convolutional layers is ever expanding, we used the layer proposed by Kipf and Welling  \cite{KipfWelling2017} designed to be analogous to the CNN setting:
\begin{equation}\label{eq_layer_graph_convolution}
H^{\left(i+1\right)} = \tilde{D}^{-\frac{1}{2}} \left( \mathcal{A} + I \right) \tilde{D}^{-\frac{1}{2}} H^{\left(i\right)} W^{(i)},
\end{equation}
where the inputs to the layer are the graph's feature matrix $H^{\left(i\right)} \in \mathbb{R}^{N\times f\left(i\right)}$ and adjacency matrix $\mathcal{A} \in \mathbb{R}^{N\times N}$, and the output is a new feature matrix $H^{\left(i+1\right)} \in \mathbb{R}^{N\times f\left(i+1\right)}$ for the graph with the same structure \cite{KipfWelling2017}. Self loops, or edges between a graph node and itself, are represented by the identity matrix $I$, of the same size as $\mathcal{A}$, and are added to the adjacency matrix. Then, that sum is multiplied on the left and right by $\tilde{D}^{-\frac{1}{2}}$ to account for the number of edges each node has. The diagonal matrix $\tilde{D}$, which does not contain trainable parameters and is only determined by $\mathcal{A}$, is defined by $\tilde{D}_{ii}=1+\sum_{j}\mathcal{A}_{ij}$. Finally, multiplication of the scaled adjacency matrix $\tilde{D}^{-\frac{1}{2}} \left( \mathcal{A} + I \right) \tilde{D}^{-\frac{1}{2}} $ by the input feature matrix $H^{(i)}$  aggregates information within local neighborhoods and multiplication by the weight matrix $W \in \mathbb{R}^{f^{\left(i\right)} \times f^{\left(i+1\right)}}$ takes linear combinations of the aggregated features to form the output features.

Bias parameters $b \in \mathbb{R}^{f^{\left(i+1\right)}}$ can be included in a graph convolution by adding them to the output feature vector of each node (each row of the output feature matrix). The weight matrix $W$ (and optional bias vector) are the trainable parameters that are optimized during training. One significant difference between convolutional layers and graph convolutional layers is in how they aggregate information within a pixel or node's neighborhood. Convolutional layers learn linear aggregation functions via the kernel parameters while graph convolutions aggregate information according to a fixed linear function, $\tilde{D}^{-\frac{1}{2}} \left( \mathcal{A} + I \right) \tilde{D}^{-\frac{1}{2}}$, determined by the graph's adjacency matrix $\mathcal{A}$. The non-learned aggregation function of such a graph convolution has raised questions of the learning capacity of graph convolutional layers \cite{Hoang2019}. Despite those concerns, graph convolutional layers have been used successfully for a variety of graph and node classification tasks \cite{Kipf2016, Zhang2019graph}.   GCNs have also been used for model-based learning directly on irregular mesh data  \cite{Herzberg2021}.  In the down-sampling path, graph convolutional layers are used like in the original graph U-net \cite{Gao2019} and in previous work on images represented as graphs \cite{Herzberg2021}. 

After graph convolutions, pooling layers are used to move down and up the U-net.  Several node-dropping, hierarchical pooling layers (layers that gradually coarsen a graph by removing nodes) for GNNs have been proposed including self-attention graph pooling \cite{Lee2019sagpool}, adaptive structure aware pooling \cite{Ranjan2019}, and gPool \cite{Gao2019}. The gPool layer selects graph nodes to preserve by first learning a projection of nodes' feature vectors. For each node, its feature vector is projected by a vector $p \in \mathbb{R}^{\left(i\right)}$ to return a scalar score. Then, the nodes with the top projection scores are preserved. The adjacency matrix is also sliced to preserve only the rows and columns for the preserved nodes. The parameters in the projection vector $p$ are optimized during training. 

One significant difference between the gPool layer and the max-pooling layer used in CNNs is that the gPool layer performs global node selection while the max-pooling layer performs local selection. That is, the max-pooling layer only considers features within a subregion or window of the input when preserving pixel data, while the gPool layer considers the projection scores from the entire graph when deciding which nodes to preserve. Self-attention graph pooling, adaptive structure aware pooling, and other node-dropping, hierarchical pooling layers also consider the whole graph when selecting which nodes to preserve, which gives way to the possibility that entire regions of a graph could be discarded when pooling \cite{Liu2022}; something that is not possible with CNN-based max-pooling.  Therefore a new graph pooling layer, the {\em k-means cluster (kMC) max pool layer}, was developed with local pooling in mind.

To create local windows in the input graph, the k-means++ algorithm\footnote{The MATLAB R2021b implementation \texttt{kmeans()} was used.} \cite{Arthur2007} was used to cluster the graph nodes. The inputs to the k-means++ algorithm include the locations of the $N$ graph nodes and the number of clusters $N_{c}$ desired. The output of the algorithm was the cluster assignments $c \in \left\{1,...,N_{c} \right\}^{N}$ and the locations of the clusters. Locations of the input nodes are determined from the FE mesh that the input graph was representing. That is, for data on the mesh nodes, the locations of those nodes are used, and, for data on the elements, the element centroid locations can be used. The locations of the output clusters are taken as the centroid of the graph nodes in the cluster.  As the k-means++ algorithm is stochastic, it was repeated multiple times, and the cluster assignments with the minimum within-cluster spacing selected.

\begin{figure}[h!]
	\centering
	\begin{picture}(240,160)
	
	\put(10,85){\includegraphics[width=0.45\textwidth]{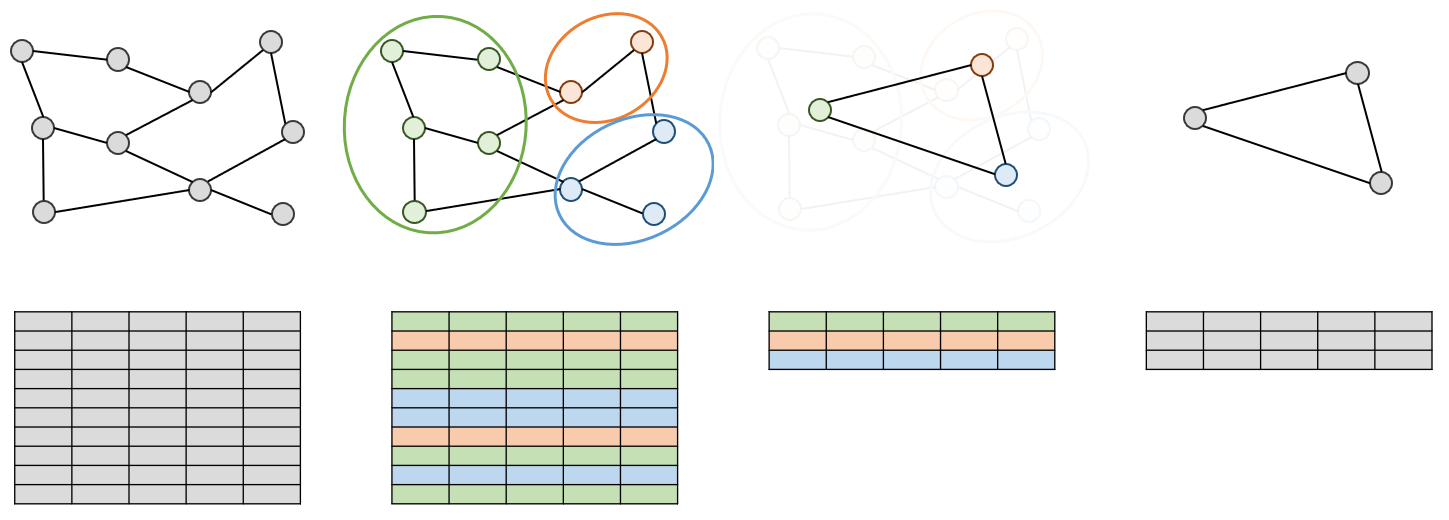}}
	\put(10, 0){\includegraphics[width=0.45\textwidth]{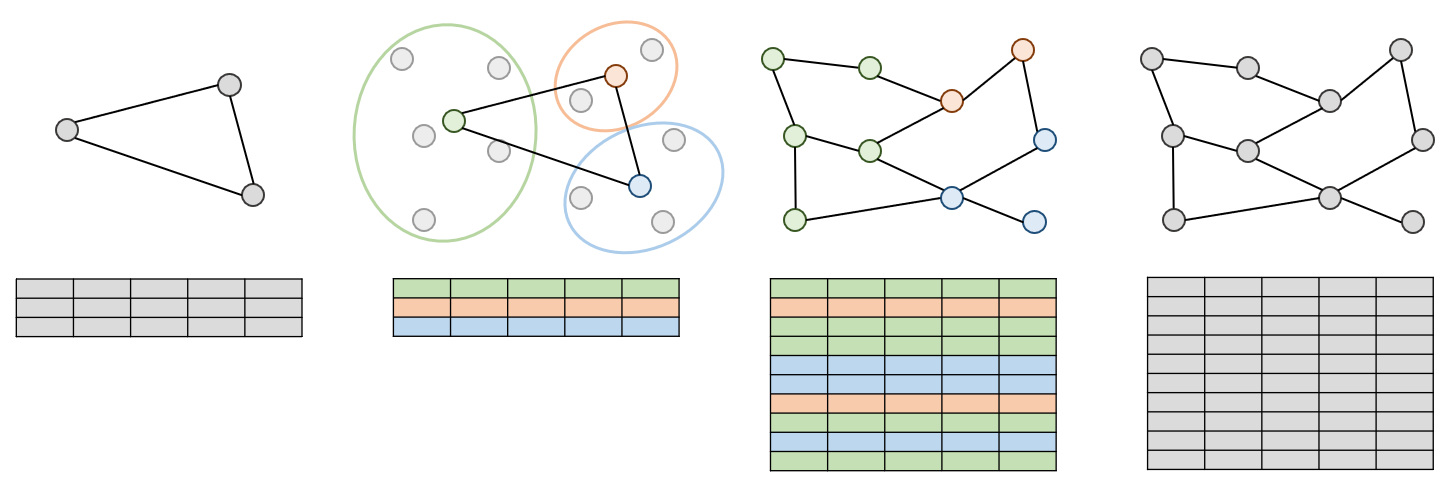}}
	
	
	\put(0,105){\rotatebox{90}{\footnotesize \em kMC max pool}}
	\put(0,5){\rotatebox{90}{\footnotesize  \em Clone Cluster unpool}}
\end{picture}

 \caption{Top: A diagram of a graph (top) and its feature matrix (bottom) being pooled using the novel \textbf{k-means cluster max pool layer}. The k-means++ algorithm is used to cluster the input nodes (left) according to their spatial location. For each cluster (middle left), an output node is placed at the centroid of the cluster (middle right), and the node's features are determined using the maximum within the cluster. The edges of the output graph (right) connect previously connected clusters.  Bottom: A diagram of a graph (top) and its feature matrix (bottom) being unpooled by a \textbf{clone cluster unpool layer} after a previous cluster-based pooling layer. The previous structure of the graph (node locations and edges) are restored (middle left) and the features of the input nodes are cloned/copied to the output nodes within each cluster (middle right).}
	\label{fig:diagrams_pooling}
\end{figure}

With clusters defined, max-pooling is performed within each cluster along each feature of the input graph. Therefore, the output of the pooling operation is a set of $N_{c}$ nodes at the cluster locations and with the maximal features from the input nodes within each cluster. The adjacency matrix of the output graph is determined by the cluster assignments as well. If any edge connected two input nodes assigned to separate clusters, an edge was drawn between the output nodes representing those clusters. Figure~\ref{fig:diagrams_pooling} (top) depicts the structure of the input and output graphs of the kMC max pooling operation.

In encoder-decoder type GNNs that have a down-sampling path of node-dropping, hierarchical pooling layers and a symmetrical up-sampling path, the up-sampling often uses the gUnpooling layer \cite{Gao2019, Liu2022}. Nodes and edges that were removed in the associated pooling layer are restored in the gUnpooling layer. The features of the restored nodes are set to zero. There are no trainable parameters in the gUnpool layer.  The gUnpool layers could not be used here as pairing it with the k-means cluster max pool layer, would result in the output graph having all nodes' features set to zero because all output nodes are restored. Like the gUnpool layer, a new {\em clone cluster unpool layer}, shown in Figure~\ref{fig:diagrams_pooling} (bottom) was designed to restore the pre-pooled graph structure. Instead of returning the previously removed nodes with features set to 0, the nodes are restored with feature vectors equal to the node representing the cluster to which the nodes belong. That is, each output node is restored as a clone/copy of the input node representing the cluster including the output node. The proposed layer was developed to act like a nearest-neighbor up-sampling layer, or a transpose convolutional layer with equal stride and kernel size and parameters fixed to~1.  

For each graph, the adjacency matrix and downsampled clustering assignments can be computed offline prior to the training or processing through the network.  The resulting graph U-net $\Lambda_\theta$ takes as inputs $\left(x_{\mbox{\tiny in}}, \mathcal{A}, \mathbf{c}\right)$ where $\mathbf{c}=\left(c_{1},...,c_{N_p}\right)$ are the  cluster assignments  for the $N_p$ pooling (and unpooling) layers in the network.  The post-processed image $\hat{x}$ is then the output of  
\begin{equation}\label{eq:post_process_gnn}
\hat{x} = \Lambda_{\theta} \left( x_{\text{in}}, \mathcal{A}, \mathbf{c} \right).
\end{equation}
Note that each cluster assignment vector, $c_{j}$, is of a different length, and $\mathbf{c}$ represents the list of vectors. Figure~\ref{fig:diagrams_unet_gnn} provides a diagram of the proposed graph U-net with $N_p=3$ pooling layers.  
The {\em k-means cluster max pooling} and {\em clone cluster unpooling} layers used in this graph U-net allow the cluster assignments to be mesh specific and computed prior to training or predicting. In the experiments conducted, no problems were noticed by training or prediction samples having different meshes and cluster assignments. Computing cluster assignments can take several minutes depending on the number of input nodes, clusters, and repetitions, thus computing them ahead of time keeps both training and prediction fast. In addition, the number of clusters used at each pooling layer can be tuned similarly to how the kernel size of a CNN max pooling layer can be adjusted. Overall, the new proposed graph pooling layer was a fast and flexible alternative to the existing graph pooling layers, and it behaves more like the traditional max pooling layer used in convolutional U-nets. The proposed clone cluster unpooling layer works naturally with the cluster-based pooling in the graph U-net architecture.  Alternatively, regional downsampling of the graph and adjacency matrix could be performed via mesh coarsening as in \cite{Suk2023}.

\subsubsection{Training}\label{sec:training}
Training data consists of $\left\{\xreci,\xtruei\right\}$ pairs where $\xreci$ is produced via the user's chosen solution method on measurement data $y^i$.  The computational mesh associated to $\xreci$ leads to an adjacency matrix $\mathcal{A}^i$ and corresponding cluster assignments $\mathbf{c}^i$.  Then, the network is trained using $\xreci$, $A^{i}$, and $\mathbf{c}^{i}$ as inputs and $\xtruei$ as the desired outputs. A loss function involving the network outputs $\hat{x}^{i}$ and the known true images $\xtruei$ can be used. Either MSE or an $\ell^1$ loss function are natural choices.  One could also weight the loss function differently for each sample or for each graph node. Once the training of the network is complete, the parameters $\hat{\theta}$ are saved for use in the online prediction stage. 

%

Note that training data could have all different computational meshes, all the same, or any mix thereof.  In this work, a single reconstruction mesh was used per network trained to demonstrate a simple case.  Note that this does not restrict later passing reconstructions on different meshes through the trained network (i.e. a network trained on reconstructions from mesh$^j$, adjacency matrix $\mathcal{A}^j$, with clusters $\mathbf{c}^j$ is not restricted only to mesh$^j$ as the adjacency matrix and clusters are inputs to the network).  Several test cases are presented in Section~\ref{sec:results} demonstrating the flexibility of the networks to input data coming from different domain shapes, experimental setups, and even higher dimensional data (3D when trained on 2D).

In the online prediction stage, an updated reconstruction $\hat{x}$ is estimated from measurement data $y$ by first computing an initial reconstruction $\xrec$ from $y$ and passing $\xrec$, its adjacency matrix $\mathcal{A}$, and corresponding cluster assignments $\mathbf{c}$ through the trained network using \eqref{eq:post_process_gnn} where the trained parameters $\hat{\theta}$, which were saved during the offline stage, are used in the network.

\subsection{Case Study: Electrical Impedance Tomography}\label{sec:eit}
EIT is an imaging modality that uses electrodes attached to the surface of a domain to inject harmless current and measure the resulting electrical potential. From the known current patterns and resulting measured potential, the conductivity distribution of the interior of the domain can be estimated \cite{Calderon1980}.  The mathematical problem of recovering the conductivity is a severely ill-posed inverse problem as large changes in the internal conductivity can present as only small changes in the boundary measurements \cite{Alessandrini1988, Nachman1996}. 

The recovered conductivity distribution can be visualized as an image and/or useful metrics extracted.  Applications of EIT are wide-ranging from nondestructive evaluation to several medical imaging applications (see \cite{Mueller2012,Borcea2002a} for a more comprehensive list).  Here we focus on absolute, also called static, EIT imaging which uses recovers the static conductivity at the time the data was collected from a single frame of experimental data.  Absolute/static imaging is important in applications such as nondestructive evaluation, breast cancer, or stroke classification where a pre-injury/illness dataset is unavailable.  Alternatively, time-difference EIT imaging recovers the change in conductivity between two frames of data.  Such time-difference data is useful in monitoring settings such are thoracic imaging of heart and lung function or stroke monitoring.  Commercial EIT systems for monitoring heart and lung function are available and used in Europe and South America.  Alternatively, frequency sweep data can be used in absolute imaging scenarios or difference imaging scenarios to further identify tissue based on the electrical properties and how they change with the frequency of the applied current.

The conductivity equation \cite{Calderon1980}
\begin{equation}\label{eq:cond_eq}
    \nabla \cdot \sigma(x) \nabla u(x) = 0 \qquad x\in\Omega\subset\R^n, \quad n=2,3,
\end{equation}
models the relationship between the electric potential $u(x)$ and conductivity $\sigma(x)$ in a domain $\sigma\subset \R^n$ with Lipschitz boundary.  In the {\em forward problem} of EIT, the voltage measurements at the electrodes are simulated for a known current pattern $T$ and bounded conductivity distribution $0< c \leq \sigma(x) \leq C < \infty$ for some constants $c$ and $C$.  Boundary conditions are given by the Complete Electrode Model (CEM) \cite{Somersalo1992} which takes into account both the shunting effect and contact impedance when modeling the electrodes.  The CEM is given by
\begin{equation}\label{eq:CEM}
\begin{array}{lclcl}
    \int_{e_{\ell}} \sigma  \frac{ \partial u }{ \partial\nu} dS &=& T_{\ell},& \quad &\ell = 1, 2, ..., L, \\
   \left. ( u + z_{\ell} \sigma \frac{ \partial u }{ \partial \nu } ) \right\vert_{e_{\ell}} &=& U_{\ell}, &\quad &\ell = 1, 2, ..., L, \\
    \left. \sigma \frac{ \partial u }{ \partial\nu } \right\vert_{\partial \Omega / \cup e_{\ell}} &= &0,& \quad &\ell = 1, 2, ..., L,    
\end{array}
\end{equation}
where $L$ denotes the number of electrodes, $e_{\ell}$ the $\ell^{\mathrm{th}}$ electrode; $z_{\ell}$, $T_{\ell}$, and $U_{\ell}$, are the contact impedance, current injected, and electric potential on the $\ell^{\mathrm{th}}$ electrode, respectively; and $\nu$ is the outward unit vector normal to the boundary. Furthermore, ensuring  $\sum\limits_{\ell=1}^{L} T_{\ell} = 0$ and $\sum\limits_{\ell=1}^{L} U_{\ell} = 0$  enforces conservation of charge and guarantees existence and uniqueness \cite{Somersalo1992, Vauhkonen1997thesis, Kaipio2000}. 

The {\em inverse problem}, determining the interior conductivity distribution $\sigma\in\Omega$ that led to the measured voltages for the known applied current patterns, was solved using the well-established Total Variation (TV) method.  The total variation of a discrete conductivity distribution 
\begin{equation}\label{eq_tv}
TV(\sigma) = \sum \abs{ \mathbf{L} \sigma },
\end{equation}
is often computed using the sparse difference matrix $\mathbf{L}$ which approximates the gradient of the conductivity distribution. It has one row $\mathbf{L}_i \in \R^{N_{M}}$ for each edge segment separating two elements in the mesh with $N_M$ elements. Each row of $\mathbf{L}$ has two nonzero elements; $d_i$ and $-d_i$ are the entries in the columns $n_i$ and $m_i$ for the $i^{\mathrm{th}}$ edge segment with length $d_i$ that separates mesh elements $n_i$ and $m_i$. TV regularized methods often use a smoothed approximation of \eqref{eq_tv} to simplify the minimization task of the absolute value term by making it differentiable.  In this work TV regularization is implemented by solving an optimization problem to obtain the iterate
\begin{equation}\label{eq:sigiter}
\sigma_{k+1}= \sigma_k + \alpha_k \left(\delta\sigma_k\right),
\end{equation}
where $\alpha_k$ is a step length, computed via a line search, that minimizes the objective function $F\left(\sigma_{k+1}\right)$ where
\begin{equation}\label{eq:TVfunctional}
F(\sigma) = \frac12\left\|U(\sigma)-V\right\|_2^2 + \lambda \sum_i\sqrt{\left(\mathbf{L}_i \sigma\right)^2 + \gamma},
\end{equation}
and the update
\begin{equation}\label{eq:update_TV}
\delta\sigma_k = - \left( J_k^T J_k + \lambda B_k \right)^{-1} \left( J_k^T \left( U_k - V \right) + \lambda B_k \sigma_k \right),
\end{equation}
where the subscript $^T$ denotes the nonconjugate transpose, $J_k = J(\sigma_k)$ is the Jacobian for iterate $\sigma_k$, $U_k= U(\sigma_k)$, and $B_k = \mathbf{L}^T E_k^{-1} \mathbf{L}$ where $E_k = \mathrm{diag}(\eta_i)$ with $\eta_i=  \sqrt{ \left( \mathbf{L}_i \sigma_{k} \right)^2 + \gamma}$

\subsection{Metrics for Success \& Experimental Setups}
Improvement in image/reconstruction quality will be assessed by several metrics  and furthermore compared to results from a classic CNN architecture.  The training data as well as various test sets (simulated and experimental) are also described here.

\subsubsection{CNN Comparison Method}\label{sec_ch4_cnn_comparison}
An alternative to the graph U-net presented in this work was to interpolate image data defined on a FE mesh to a pixel grid so that a typical convolutional U-net can be used.  Thus, the reconstruction $\xin$ was computed, as before, on a FE mesh from measurement data $y$, then interpolated to a pixel grid $\tilde{x}_{\text{in}} = f\left(\xin \right)$.  Next, a post-processing CNN $\Lambda_{\theta}^{\mbox{\tiny CNN}}$ is used to estimate the reconstruction on the pixel grid: $\tilde{x} = \Lambda_{\theta} \left( \tilde{x}_{\text{in}} \right)$.  If desired, the network output can then be interpolated back to the FE computational mesh: $\hat{x} = f^{\dagger} \left( \tilde{x} \right).$

In this setting, the CNN networks are trained using a set of initial reconstruction and truth image pairs, where both have been interpolated from the computational mesh  to the pixel grid. 
It may be desirable to utilize a CNN to post-process images, as opposed to a GNN, as there is  precedent for using CNNs on image data. Still, the loss of fine detail in refined portions of the mesh and the errors induced by interpolating to and from a pixel grid could be prohibitive or time intensive in certain imaging applications. 

%

\subsubsection{Experimental Data}\label{sec:experiments}
\begin{figure}[h!]
	\centering
	\begin{picture}(240,50)
	

	\put(  0,  0){\includegraphics[width=55pt]{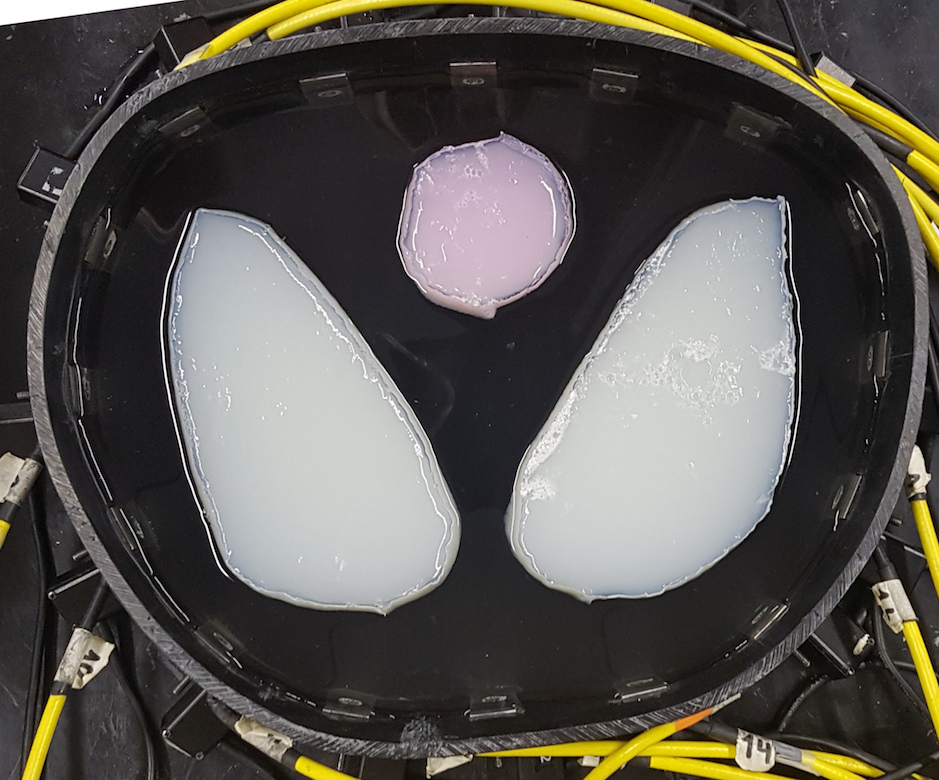}}
	\put( 60,  0){\includegraphics[width=55pt]{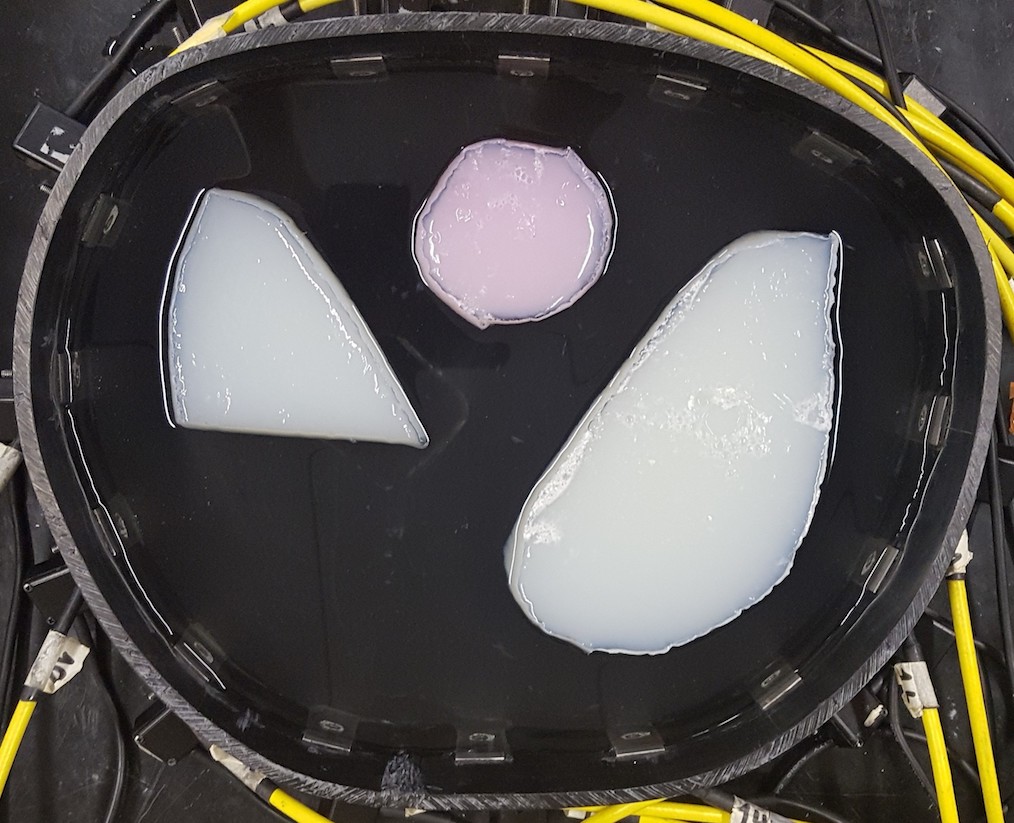}}
	\put(120,  0){\includegraphics[width=55pt]{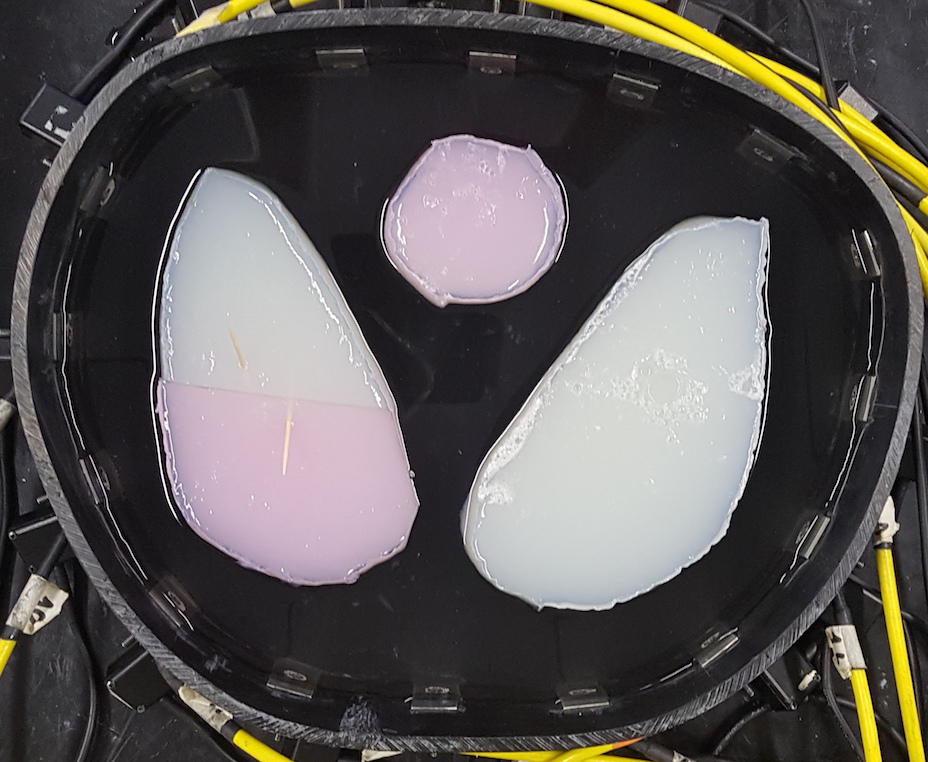}}
	\put(180,  0){\includegraphics[width=50pt]{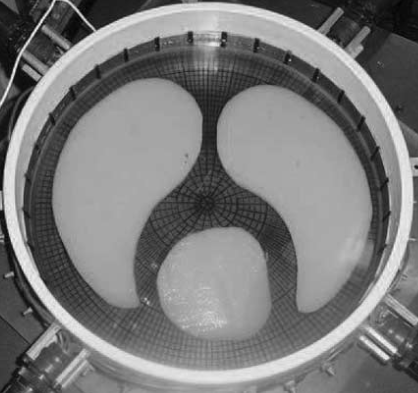}}
	
	\put( 14, 50){\underline{\footnotesize \textsc{KIT4 S.1}}}
	\put(73, 50){\underline{\footnotesize\textsc{KIT4 S.2}}}
	\put(132, 50){\underline{\footnotesize\textsc{KIT4 S.3}}}
	\put(188, 50){\underline{\footnotesize\textsc{ACT3 S.1}}}
	
	\end{picture}
	\caption[Experimental Dataset Setups]{Photographs of the KIT4 and ACT3 experimental data \cite{Isaacson2004, Hamilton2019}.}

	\label{fig:2DExperimental}
\end{figure}
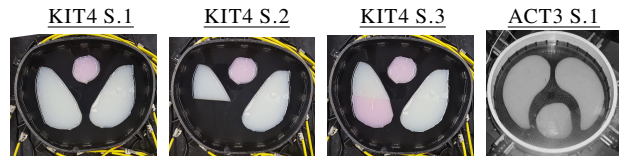
Six experimental tank datasets with conductive agar targets, taken on three different EIT machines will be used to evaluate the graph U-net method presented in Section~\ref{sec:gnns}.  The first set, denoted KIT4, comes from the 16 electrode KIT4 system at the University of Eastern Finland \cite{Kournunen2008, Hamilton2019beltrami}.  Conductive agar targets (pink 0.323~S/m, white 0.061~S/m) were placed in a saline bath, with measured conductivity 0.135~S/m, in a chest shaped tank with perimeter 1.02~m and 16 evenly spaced electrodes of width 20~mm.  The computational reconstruction mesh used for the KIT4 datasets contained 3,984 elements.  Three experiments were performed to mimic heart (pink) and lung (white) imaging (Fig.~\ref{fig:2DExperimental}-left).   Sample KIT4-S.1 shows a `healthy' setup with two low conductivity targets (lungs) and one high conductivity target (heart).  Sample KIT4-S.2 has a cut in the ``lung'' target on the viewer's left while KIT4-S3 replaces the missing portion with a more conductive agar (e.g. possibly a pleural effusion).  For each setup, 16 adjacent current patterns were applied with an amplitude of 3~mA and current frequency of 10~kHz, and measurements were recorded on all electrodes.   The regularization parameters for TV were selected as $\lambda=5\cdot 10^{-5}$ and $\gamma = 10^{-14}$ based on  testing from simulated datasets. 

Next, archival data from the 32~electrode ACT3 system \cite{Cook1994,Isaacson2004} was used.  In Sample ACT3-S.1 ((Fig.~\ref{fig:2DExperimental}-right), a circular heart target (0.750~S/m) and two lung targets (0.240~S/m) were placed in a saline bath (0.424~S/m) in a tank of radius 0.15m.  Trigonometric current patterns with maximum amplitude 0.2mA and frequency 28.8kHz were applied on the 32 equally spaced electrodes of width 25mm.  

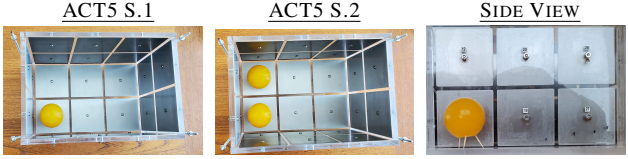
\begin{figure}
    	\centering
	\begin{picture}(240,50)
    
    \put(0,0){\includegraphics[width=75pt]{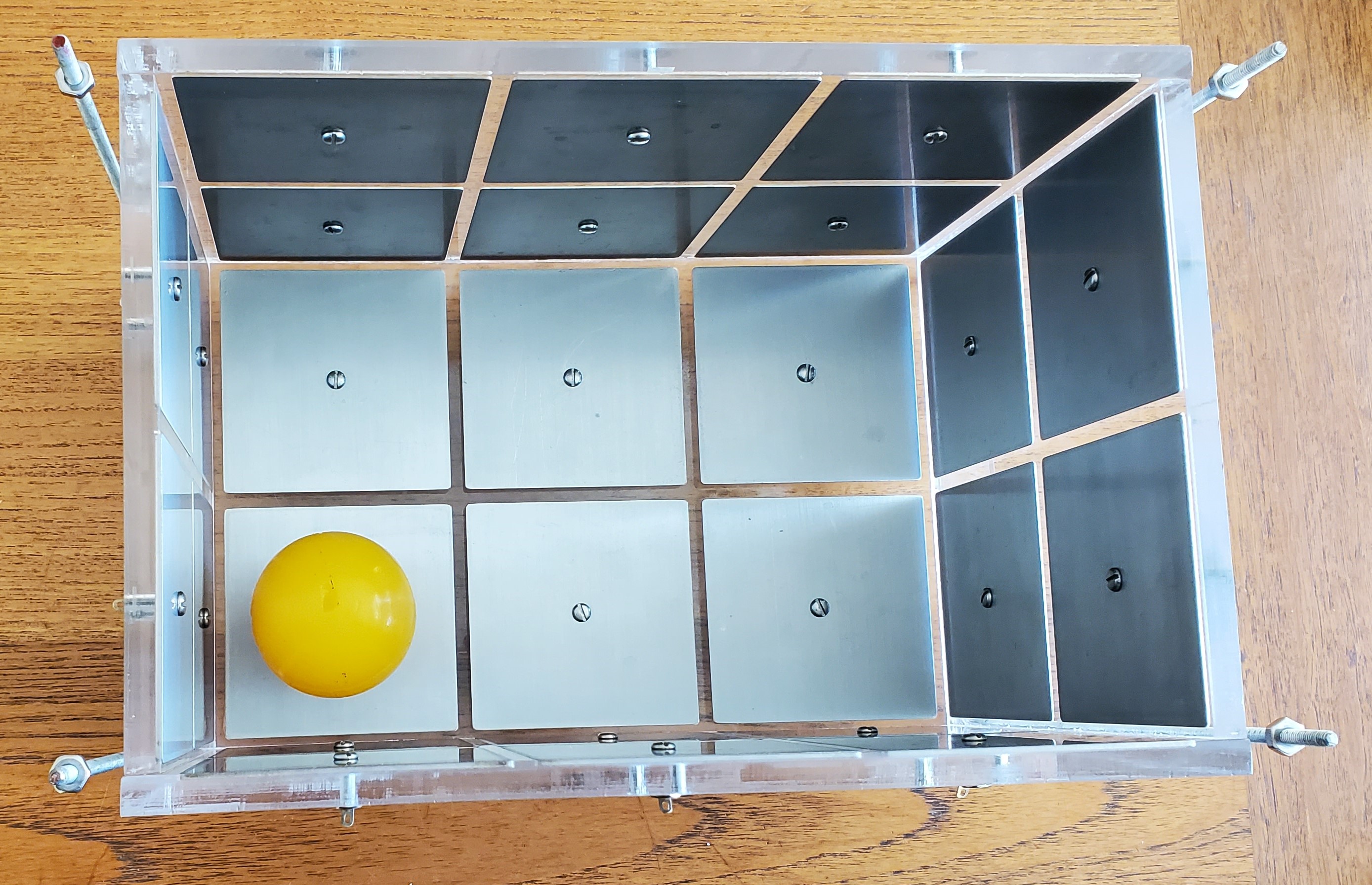}}
    \put(80,0){\includegraphics[width=75pt]{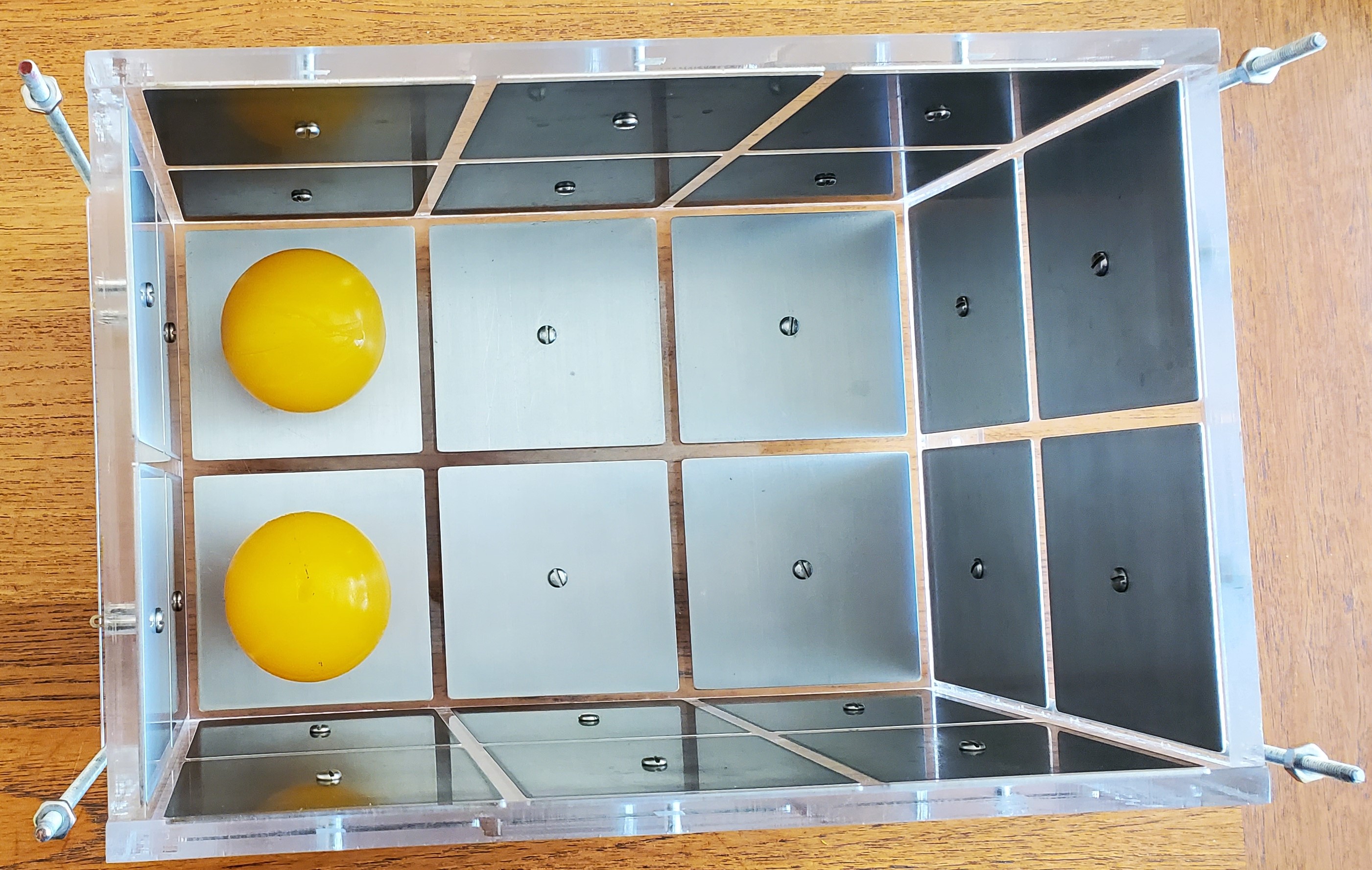}}   
     \put(160,0){\includegraphics[width=75pt]{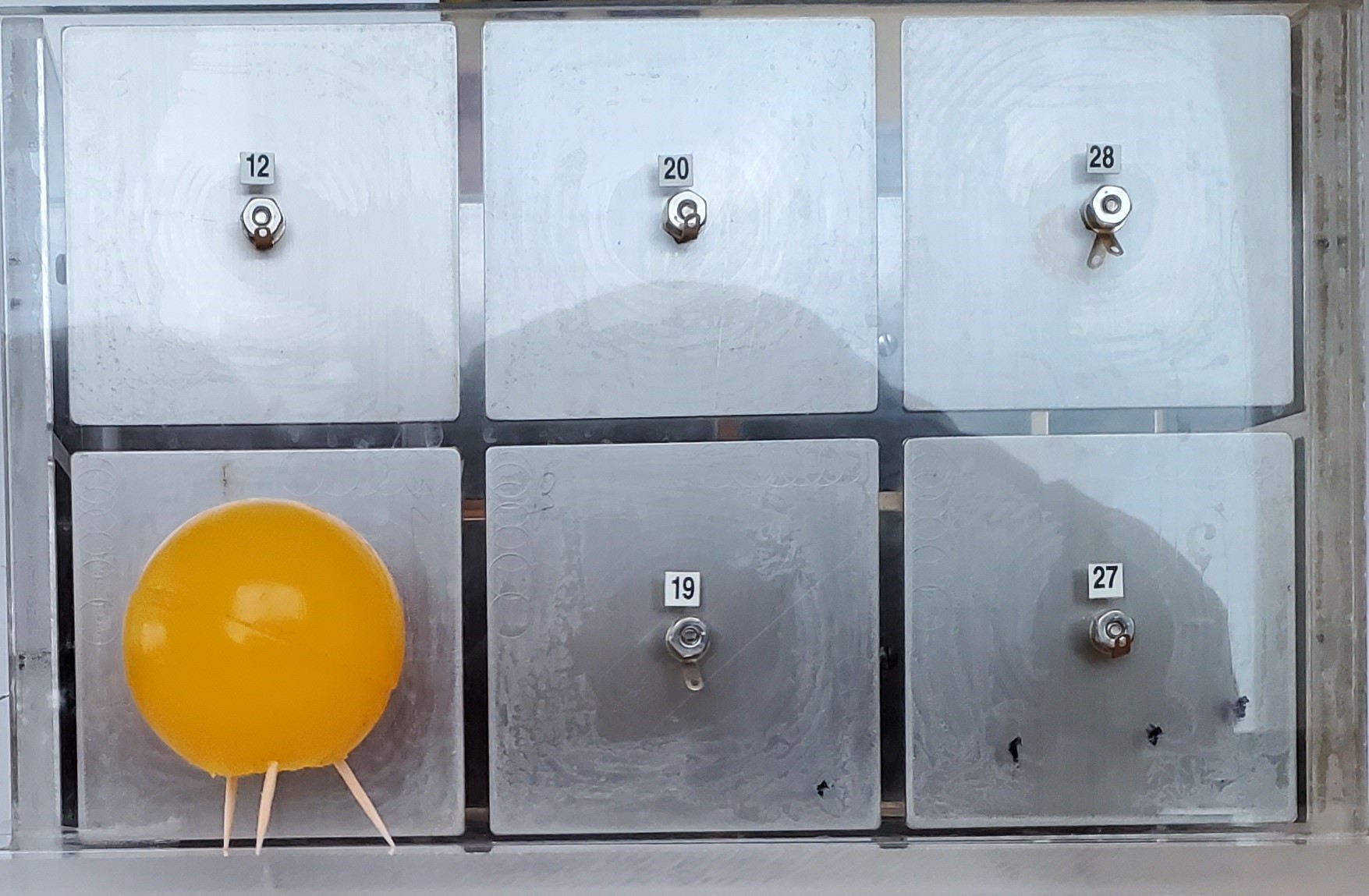}}
    
	\put( 22, 52){\underline{\footnotesize \textsc{ACT5 S.1}}}
	\put(100, 52){\underline{\footnotesize\textsc{ACT5 S.2}}}
	\put(180, 52){\underline{\footnotesize\textsc{Side View}}}
    \end{picture}
    \caption{ACT5 experimental setups and side view showing target height.}
    \label{fig:ACT5exper}
\end{figure}

Lastly, data from the ACT5 system \cite{rajabishishvan2021}, was used for extension testing to 3D data.  Samples ACT5-S.1 and ACT5-S.2 (Fig~\ref{fig:ACT5exper}) were collected on plexiglas box with interior dimensions 17.0cm~x~25.5cm x~17.0cm with 32 electrodes of size 8cm~x~8cm \cite{Hamilton2022}.  The top of the tank is removable and has small holes allowing for filling and resealing between experiments.  Spherical agar targets of measured conductivity 0.290~S/m were placed in tap water measuring 0.024~S/m.  Optimal current patterns for the saline only tank were obtained and used for ACT5-S.1 and S.2.\footnote{The experimental ACT5 data is freely available at \url{https://github.com/sarahjhamilton/open3D_EIT_data}.}

\subsubsection{Training Data}\label{sec:trainingData}
Simulated data using the chest-shaped domain corresponding the the KIT4 data, with 16 electrodes and adjacent current pattern injection, was used to generate the training data. The measurement mesh contained 10,274 elements, and the reconstruction mesh was the same as the one used for the experimental KIT4 datasets (3,984 elements). The simulated true conductivity distributions had an equal chance of either three or four randomly placed elliptical targets that were not allowed to overlap or touch the boundary. For each ellipse, the major axis was between [0.03-0.07] meters and the minor axis was [50 - 90]\% of the value of the major axis.  Each target had an equal chance of having a constant conductivity in the range of $\left[0.04, 0.07 \right]$~S/m or $\left[0.25, 0.35 \right]$~S/m, while the background conductivity values were constant in the range of $\left[0.11, 0.17 \right]$~S/m. The measured voltage data was simulated using all 16 possible adjacent current patterns with an amplitude of 3~mA and 0.5\% relative noise added to the voltages prior to reconstruction with TV.

Eight separate U-nets were trained, four based on GNNs and four on CNNs.  The four graph U-nets are named GNN-TVx and the four convolutional U-nets are named CNN-TVx where the ``x'' represents the TV method iterate used as input to the network. That is GNN-TV2 and CNN-TV2 are a graph U-net and classic U-net, respectively, that use the second iteration $\sigma_2$ of the TV method as input.  The networks were trained using 5,000 training samples and 500 validation samples. The number of training samples used was greater than the number used for the model-based methods in \cite{Herzberg2021} since the networks have many more trainable parameters. The number of samples was in line with other implementations of the U-net architecture for EIT \cite{Hamilton2019beltrami} and did not result in severe or harmful over-fitting of the networks trained here.  More testing could be done to determine if even fewer samples could be used for training. The ADAM optimizer \cite{Kingma2015} with an initial learning rate of $5\cdot 10^{-4}$ and mini batches of 32 samples were used to optimize the parameters. Other learning rates and batch sizes were also tested and resulted in similar minimum loss values and training times. Training of each network was stopped if the validation loss, MSE, failed to decrease over the course of 50 epochs and the parameters (weights and biases) that resulted in the lowest validation loss were saved. 

\begin{figure}
    \centering
	\begin{picture}(240,125)
	
	\put( 10,70){\includegraphics[width=55pt]{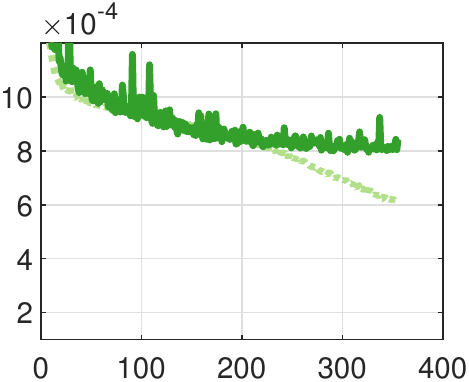}}
	\put(70,70){\includegraphics[width=55pt]{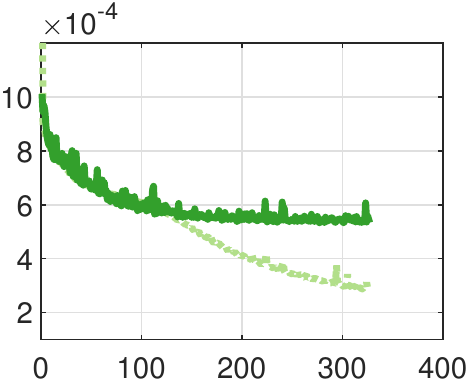}}
	\put(130,70){\includegraphics[width=55pt]{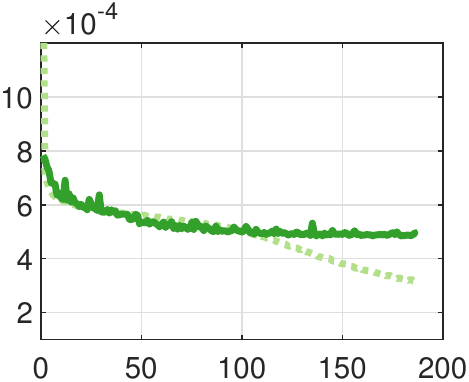}}
	\put(190,70){\includegraphics[width=55pt]{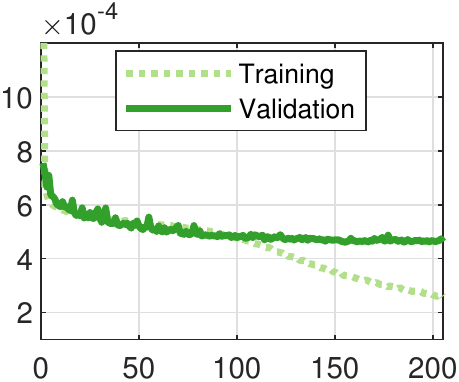}}
	
	\put( 10, 10){\includegraphics[width=55pt]{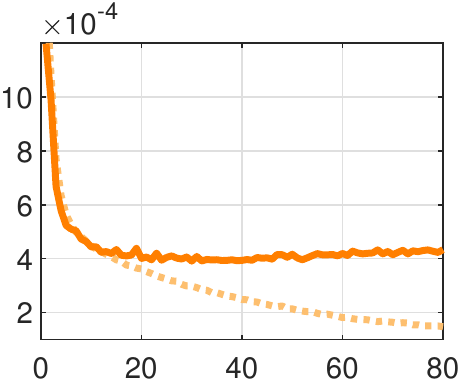}}
	\put(70, 10){\includegraphics[width=55pt]{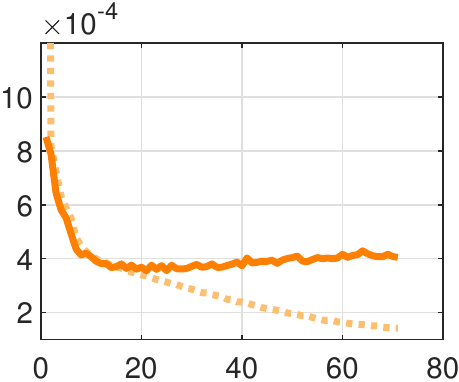}}
	\put(130, 10){\includegraphics[width=55pt]{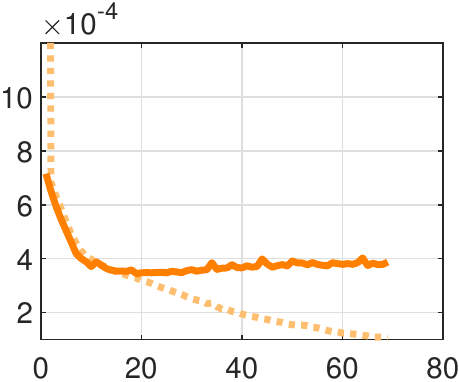}}
	\put(190, 10){\includegraphics[width=55pt]{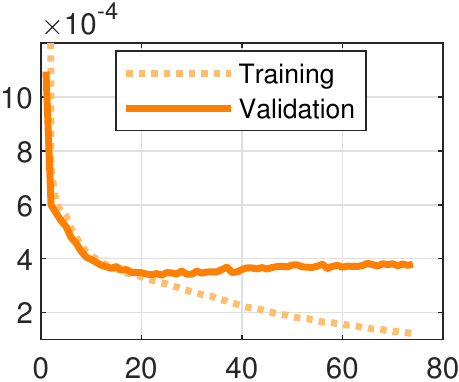}}
	
	\put(115,  0){\scriptsize\underline{\textsc{Epoch}}}
	\put(  0, 52){\rotatebox{90}{\scriptsize\underline{\textsc{Loss}}}}
	
	\put(25,120){\tiny \underline{\textsc{ GNN-TV1}}}
	\put(85,120){\tiny \underline{\textsc{ GNN-TV2}}}
	\put(145,120){\tiny \underline{\textsc{ GNN-TV3}}}
	\put(205,120){\tiny \underline{\textsc{ GNN-TV4}}}
	
	\put( 25, 60){\tiny \underline{\textsc{CNN-TV1}}}
	\put(85, 60){\tiny \underline{\textsc{CNN-TV2}}}
	\put(145, 60){\tiny \underline{\textsc{CNN-TV3}}}
	\put(205, 60){\tiny \underline{\textsc{CNN-TV4}}}
	
	\end{picture}
	\caption[GNN and CNN Loss Plots]{Loss plots (training and validation) for the \textcolor{OliveGreen}{\textbf{GNN-TVx}} and \textcolor{Orange}{\textbf{CNN-TVx}} networks that were trained with the \textbf{Chest-16} dataset. The losses were computed as the mean squared error between the networks' outputs and the true conductivity distributions computed over the mesh for the GNNs (top) and over the pixel grid for the CNNs (bottom).} \label{fig:matlab_sec_4_2_1_loss}
\end{figure}
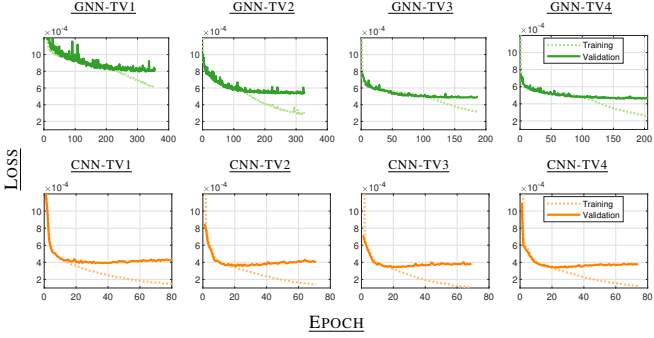

The loss plots for the GNNs and CNNs using different iterates as input are shown in Figure~\ref{fig:matlab_sec_4_2_1_loss}. As expected, the U-nets using later iterates as input achieved lower minimum validation losses than the networks using TV1 inputs. There was a greater difference in the minimum validation loss values between graph U-nets with different inputs compared to classic U-nets with different inputs. Still, for both types, there was not a large difference between the U-nets using the third and fourth iterates as inputs. The CNNs reached lower raw minimum validation loss values but they cannot be compared directly to the graph U-nets because the loss was computed over different domains and discretizations.

The shapes of the loss curves are different between the types of U-nets but consistent across input iteration. The graph U-nets required 130-280 epochs to reach a minimum validation loss while the convolutional U-nets required only 20-30 epochs. After reaching the minimum validation loss, the graph U-nets validation loss values leveled out in later epochs, while the CNN validation loss values slightly increased. For both network types, the training loss values continued to decrease. All of these characteristics were consistent across repetitions of training independent networks, and more research is needed to determine why the differences exist or what the effects are in the final reconstructions. In addition, determining if the network weights resulting in the lowest validation loss produce the ``best'' reconstructions is also the topic of future research as metrics other than MSE are also critical in assessing EIT reconstruction quality.


\subsubsection{Metrics}\label{sec:metrics}
As there is no universally accepted metric for assessing the quality of EIT reconstructions, the metrics listed below, along with visual inspection, will be used collectively to assess reconstruction quality.

\noindent\;\;\;\textbf{Mean Squared Error}: $\mathrm{MSE} = \frac{1}{N_{\scriptscriptstyle M}} \sum_{i=1}^{N_{\scriptscriptstyle M}} \left( \sigma_{\text{true},i} - \hat{\sigma}_i \right)^{2}$.

\noindent\;\;\;\textbf{Relative $l_1$ Conductivity Error}: $\mathrm{RE}_\sigma^{l_1} = \frac{ \norm{ \hat{\sigma} - \sigma_{\text{true}} }_1 }{ \norm{\sigma_{\text{true}}}_1 }$.

\noindent\;\;\;\textbf{Dynamic Range}: $\mathrm{DR} = \frac{\mathrm{max}(\hat{\sigma})-\mathrm{min}(\hat{\sigma})}{\mathrm{max}(\sigma_{\text{true}})-\mathrm{min}(\sigma_{\text{true}})} \times 100\%.$

\noindent\;\;\;\textbf{Total Variation Ratio}: $\mathrm{TVR} = \frac{\sum \abs{ \mathbf{L} \hat{\sigma} }}{\sum \abs{ \mathbf{L} \sigma_{\text{true}} }} \times 100\%$.

\noindent\;\;\;\textbf{Relative $l_2$ Voltage Error}: $\mathrm{RE}_V^{l_2} = \frac{ \norm{ U(\hat{\sigma}) - V }_2 }{ \norm{ V }_2 }.$

Note that these metrics above do not account for the varying sizes of the mesh elements.  If desired, the metrics can be scaled relative to element size as well and the networks even trained based on such weighted metrics.  Further work is needed to determine if the weighted or unweighted metrics and/or using weighted loss functions during training are more correlated with visually high quality reconstructions. Such work, while interesting is left for future studies.  Here, only unweighted metrics were used and reported.  Additionally, region of interest (ROI) metrics may be of higher interest than global image metrics.  Where appropriate, e.g. lung imaging, ROI metrics are also presented.

\section{Results}\label{sec:results}

\begin{figure}
\linethickness{.3mm}
    \centering
\includegraphics[width=0.5\textwidth]{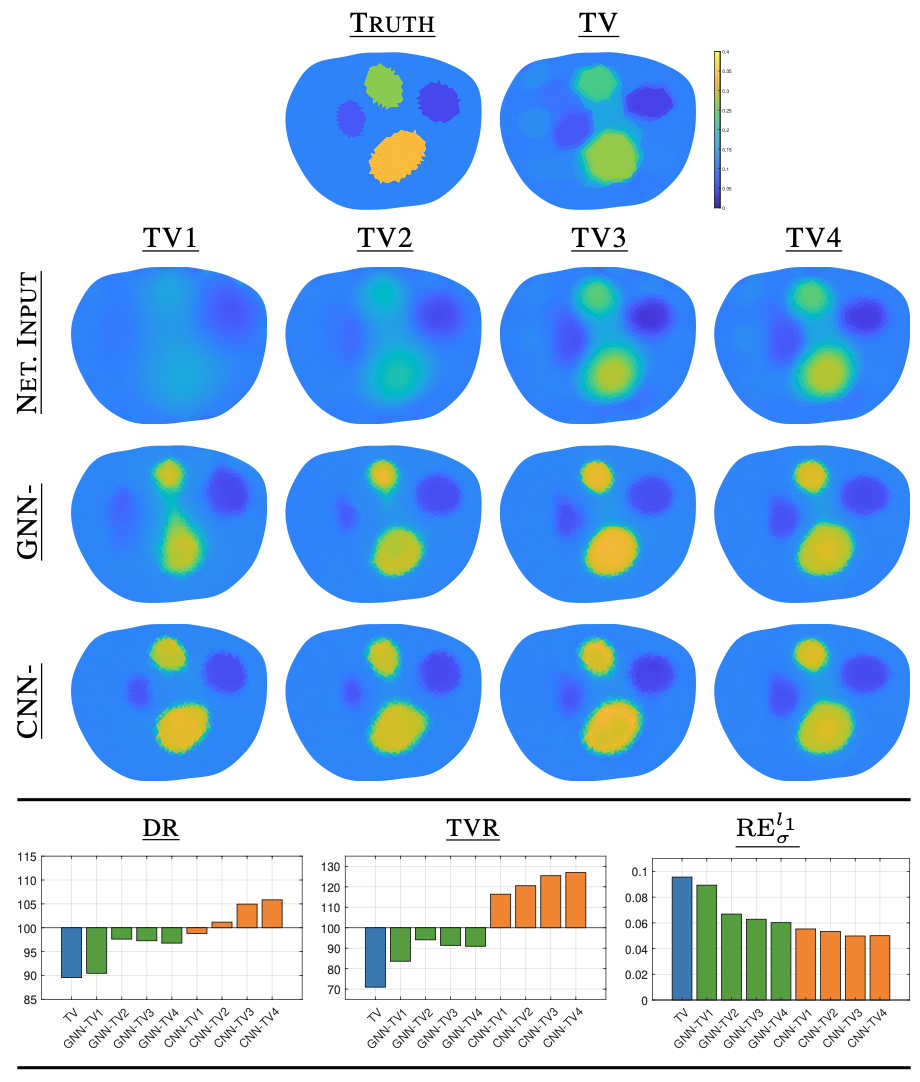}
	\caption{Demonstration of the \textcolor{OliveGreen}{\textbf{GNN-TVx}} and \textcolor{Orange}{\textbf{CNN-TVx}} networks on simulated data consistent with training data as well as average metric scores for reconstructions of such 50 test samples compared to the \textcolor{RoyalBlue}{\textbf{TV}} method.  All conductivity reconstructions are on the same color scale.}
    \label{fig:fig4o8}
\end{figure}

\begin{figure*}[bt]
    \centering   
    \includegraphics[width=\textwidth]{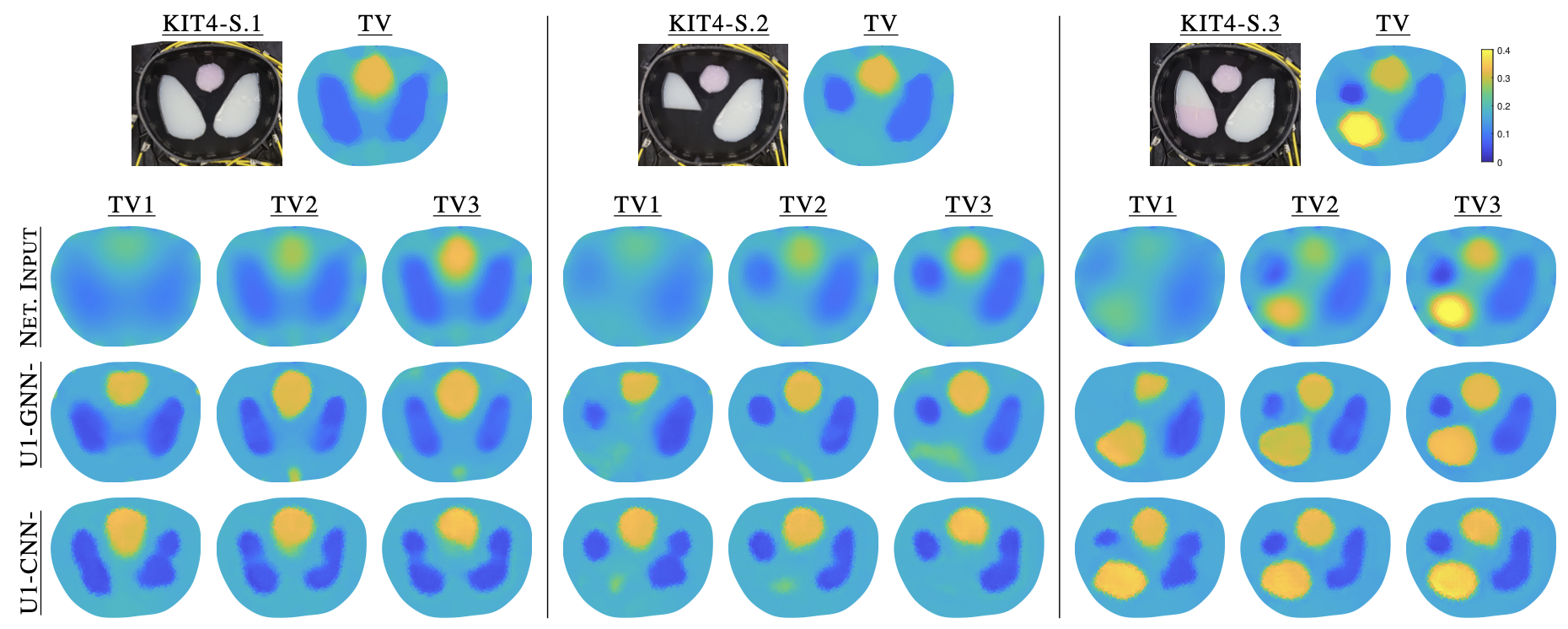}

	\caption{Results for KIT4-S.1, S.2, and S.3 for the first three iterations of TV as input to the networks GNN-TVx and CNN-TVx, all on the same color scale.}
    \label{fig:KIT4recons}
\end{figure*}

We first explore results for data consistent with the training data described in Sec.~\ref{sec:trainingData}.  Figure~\ref{fig:fig4o8} compares the results for a simulated dataset consistent with the network data.  The top row displays the truth as well as the non-learned TV reconstruction after 20 iterations.  The second row displays the input images for the networks which vary from iteration 1 to 4 of TV.  The third and fourth rows show the post-processed reconstructions from the GNN and CNN networks.  Note that the inputs (row 2) are interpolated to a pixel grid, processed by the CNN networks, then interpolated back to the computational mesh on which they are displayed in row four.  As expected the CNNs display excellent sharpening even from a first TV iteration input image.  The GNNs outputs are similarly sharpened with early iterate TV inputs but the targets are better separated when using at least the second TV iterate.  
Metrics averaged over 50 simulated test samples consistent with the training data are also shown in Fig~\ref{fig:fig4o8}. We see that overall the learned methods outperform the classic TV in all metrics aside from the relative voltage error fit, which is not unexpected as TV optimizes specifically for this whereas the network solutions do not.  The GNN and CNN based U-nets perform similarly across the remaining metrics with the GNNs slightly outperforming the CNNs for the DR and TVR.  Note, also that the first iteration of TV performed significantly worse than later TV iterates for the GNNs indicating a second iterate starting point may be advisable.

\begin{table}[]
\scriptsize
\centering
\setlength{\tabcolsep}{2pt}
\caption{ROI metrics (S/m) for KIT4 and ACT3 reconstructions.}
\begin{tabular}{r|l|c|c||c|c|c|c||c|c|c|c|}
\hline
\hline
    &         & \multirow{2}{*}{Truth} & \multirow{2}{*}{TV} & \multicolumn{4}{c}{GNN-}            & \multicolumn{4}{c}{CNN-}            \\
      &         &                        &                     & TV1            & TV2   & TV3   & TV4   & TV1   & TV2   & TV3   & TV4            \\

\hline
\\
\multirow{3}[0]{*}{\rotatebox{90}{KIT4-S.1} }& Heart   & 0.323                  & \textbf{0.324}      & 0.309          & 0.319 & 0.326 & 0.321 & 0.325 & 0.326 & 0.336 & 0.332          \\
                          & R. Lung & 0.061                  & 0.105               & \textbf{0.100} & 0.116 & 0.118 & 0.118 & 0.106 & 0.113 & 0.104 & 0.106          \\
                          & L. Lung & 0.061                  & 0.099               & 0.101          & 0.109 & 0.110 & 0.111 & 0.098 & 0.104 & 0.106 & \textbf{0.099} \\
                          \\
                          \hline
                          \\
\multirow{3}[0]{*}{\rotatebox{90}{KIT4-S.2} } & Heart   & 0.323                  & 0.318               & 0.291          & 0.317 & \textbf{0.324} & \textbf{0.322} & 0.326 & \textbf{0.324} & 0.334 & 0.327          \\
                          & R. Lung & 0.061                  & 0.105               & \textbf{0.098}          & 0.112 & 0.116 & 0.120 & 0.106 & 0.114 & 0.105 & 0.108          \\
                          & L. Lung & 0.061                  & \textbf{0.101}               & 0.122          & 0.105 & 0.107 & 0.103 & 0.109 & 0.105 & 0.109 & \textbf{0.101}          \\
                          
\\                          
\hline
\\
\multirow{3}[0]{*}{\rotatebox{90}{KIT4-S.3} } & Heart   & 0.323                  & 0.299               & 0.271          & 0.309 & 0.320 & 0.318 & 0.320 &\textbf{0.323} & 0.327 & 0.320          \\
                          & R. Lung & 0.061                  & 0.099               & \textbf{0.097}          & 0.105 & 0.103 & 0.106 & 0.106 & 0.106 & 0.100 & 0.105          \\
                          & L. Lung & 0.061                  & \textbf{0.118}               & 0.176          & 0.136 & 0.125 & 0.144 & 0.141 & 0.128 & 0.123 & 0.120          \\
                          & Injury  & 0.323                  & 0.350               & 0.302          & 0.295 & \textbf{0.320} & 0.316 & 0.337 & 0.328 & 0.344 & 0.357         \\
                          \\
                          \hline
                          \hline
                           \\
\multirow{3}[0]{*}{\rotatebox{90}{ACT3-S.1} } & Heart   & 0.750                  & \textbf{0.748}              & 0.434          & 0.780 &0.788 & 0.851 & 0.426 & 0.656 &0.765 & 0.808         \\
                          & R. Lung & 0240                  & 0.216               & 0.277         & 0.200& 0.225 & 0.217& 0.292& \textbf{0.231} & 0.204 & 0.207          \\
                          & L. Lung & 0.240                  & 0.202               & 0.271          & 0.205 & 0.223& 0.215 & 0.280 & \textbf{0.230}& 0.196 & 0.200         \\
                          \\
                          \hline
                          \hline

\end{tabular}
\label{tab:ROIs}
\end{table}

Reconstructions from the experimental KIT4 S.1-S-3 datasets are shown in Fig.~\ref{fig:KIT4recons}, with corresponding ROI metrics are shown in Table~\ref{tab:ROIs}.  As with the simulated data, we see remarkable sharpening even with a single iterate of TV used as input for both the CNN and GNN.  Recall that the training data for the networks was simple ellipses and thus the target shapes even in KIT4-S.1 are slightly different than that and the networks have not seen `cut' data as in KIT4-S.2 and S.3.  The post-processed images from the CNNs have slightly deformed `lungs' when compared to the GNN output.  Small conductive artifacts appear many of the GNN and CNN reconstructions at the bottom center.  None of the methods, learned or TV, were able to recover the sharp cut in both the top and bottom portion of the viewer's left lung in Sample KIT4-S.3.  The bottom portion of the lung did have a sharper dividing line for the GNNs as well as CNN-TV4.  Moving to the metrics, we see that MSE and $\mathrm{RE}_\sigma^{l_1}$ are even, TV slightly outperforms the networks as expected.  However, in the DR and TVR we again see the GNNs better approximate the true dynamic range and total variation ratio, in particular for the earlier TV iterates.  For the ROI metrics in Table~\ref{tab:ROIs}, overall, the GNNs slightly outperform both the full TV and CNNs.  In each case the full TV reconstruction produced the visually most similar reconstruction to the truth, but needed approximately 20 iterations, compared to 1-4 iterations with the post-processed setting.  Depending on the application, in particular for 3D, it may be important to balance the reconstruction quality versus computational cost.

\begin{figure}
    \centering
    \includegraphics[width=0.5\textwidth]{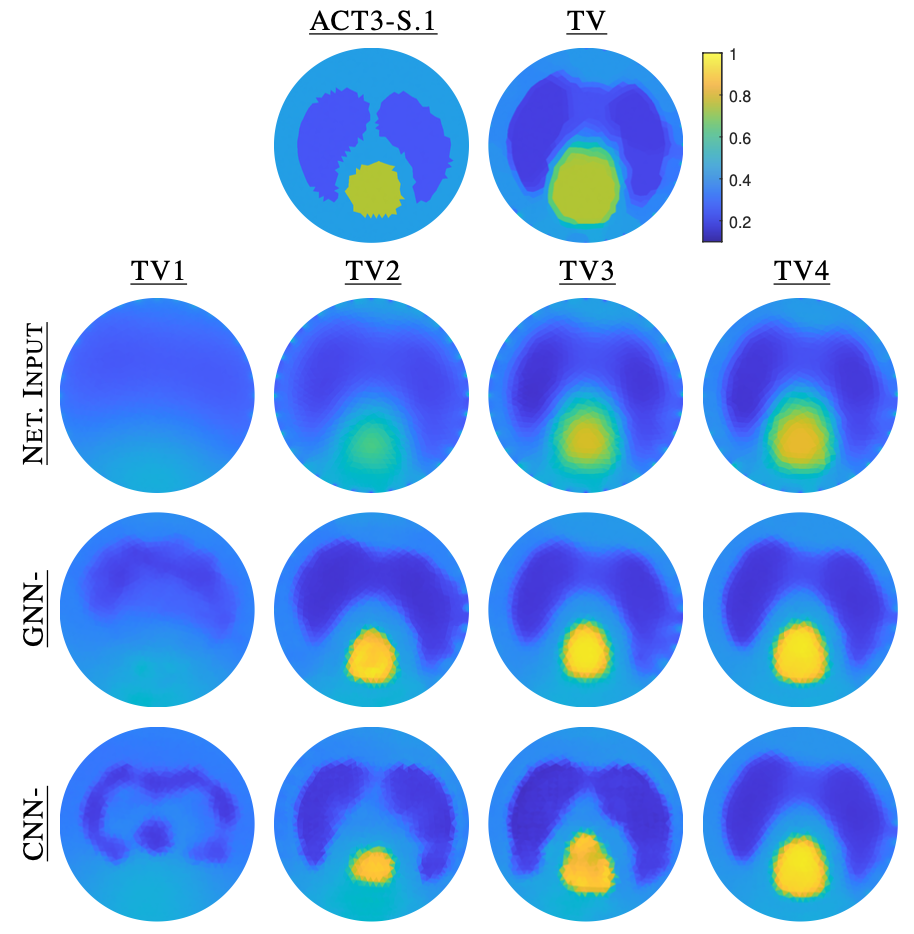}
	\caption{Demonstration of the networks on the ACT3-S.1 data.}
    \label{fig:ACT3recons}
\end{figure}

\begin{figure}
\linethickness{.3mm}
    \centering
        \includegraphics[width=0.5\textwidth]{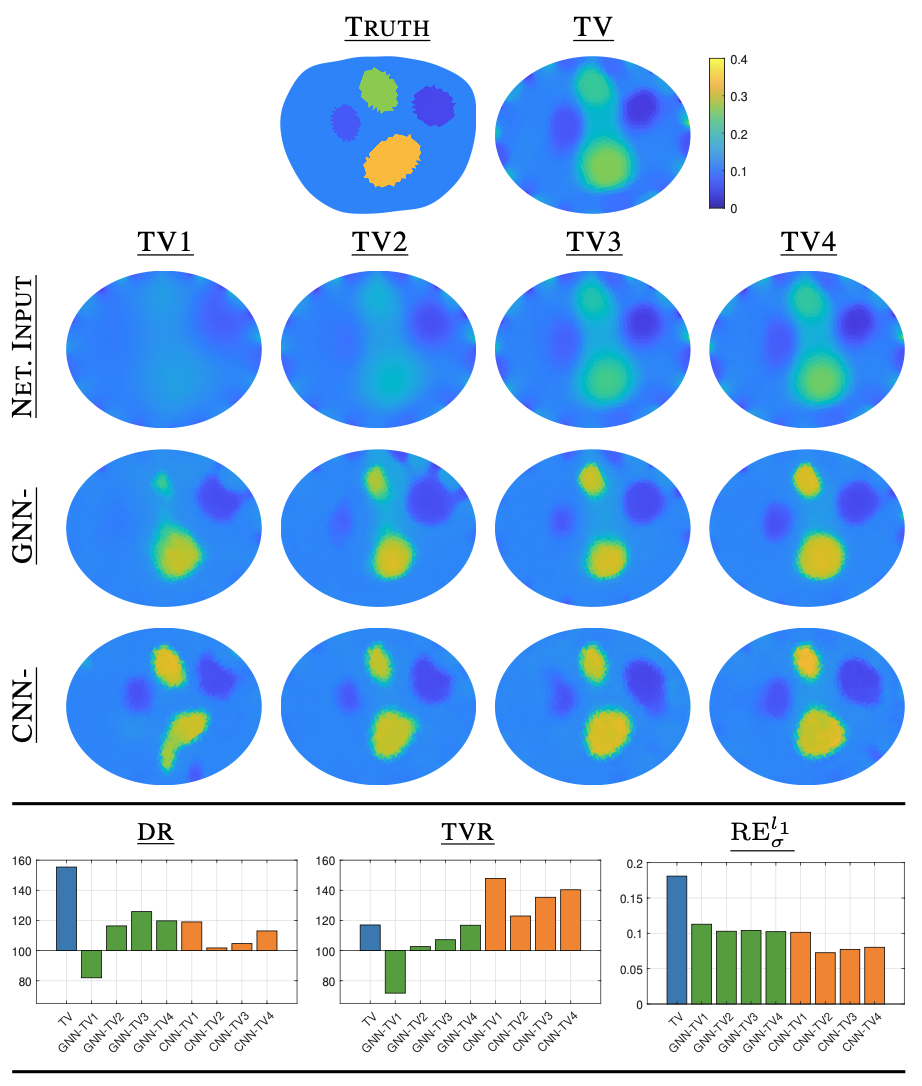}
	\caption{Robustness to domain modeling errors.  Results of the \textcolor{OliveGreen}{\textbf{GNN-TVx}} and \textcolor{Orange}{\textbf{CNN-TVx}} networks on simulated data, assuming the true domain was an ellipse instead of chest-shape as well as average metric scores for reconstructions of such 50 test samples compared to the \textcolor{RoyalBlue}{\textbf{TV}} methods.}
    \label{fig:ellipse}
\end{figure}

Next, further out of distribution data is tested by using the ACT3-S.1 dataset which comes from a circular domain, 32 electrodes, applied trigonometric current patterns, and has targets with different conductivity values than in the training data.  In order to process the ACT3-S.1 sample through the trained networks, the input images were scaled to have the background value in the expected window, processed, then scaled back.  Figure~\ref{fig:ACT3recons} shows the resulting conductivity reconstructions and Table~\ref{tab:ROIs} the corresponding ROI metrics.  Here, both the CNN and GNN networks required at least TV2 input to resolve the targets, though the GNN-TV1 image is more recognizable than the CNN-TV1, albeit the contrast is worse.  The CNN-TV2 did the best at separating the lungs but underestimated the size of the heart.  The ROI metrics show that the mean target ROIs are quite close overall to the truth.

To this end, in  Fig.~\ref{fig:ellipse}, we study how well the networks handle incorrect domain modeling, a notoriously challenging problem in absolute/static EIT imaging \cite{Hamilton2018robust}.  Here, the true domain is the chest-shaped domain but we naively reconstruct on the elliptical domain shown.  Again, we see that TV1 is not informative enough for GNN-TV1 to recover all four targets.  Notably, the boundary artifacts common from the domain mismodeling are significantly reduced in the post-processed images for both the GNN and CNN networks.


%

\section{Discussion}\label{sec:discussion}
To further test test the robustness of the graph U-net we explored how well the networks trained on 2D EIT data worked when the input data came from 3D measurement data that was significantly different than the networks were trained on.  
Additional modifications to the network structure (layers and inputs) are discussed.  Lastly, a free python package GN4IP developed for this project is presented.

\subsection{Testing 3D data in 2D networks}\label{sec:3Din2D}
\begin{figure}
    	\centering
	
	    \includegraphics[width=0.5\textwidth]{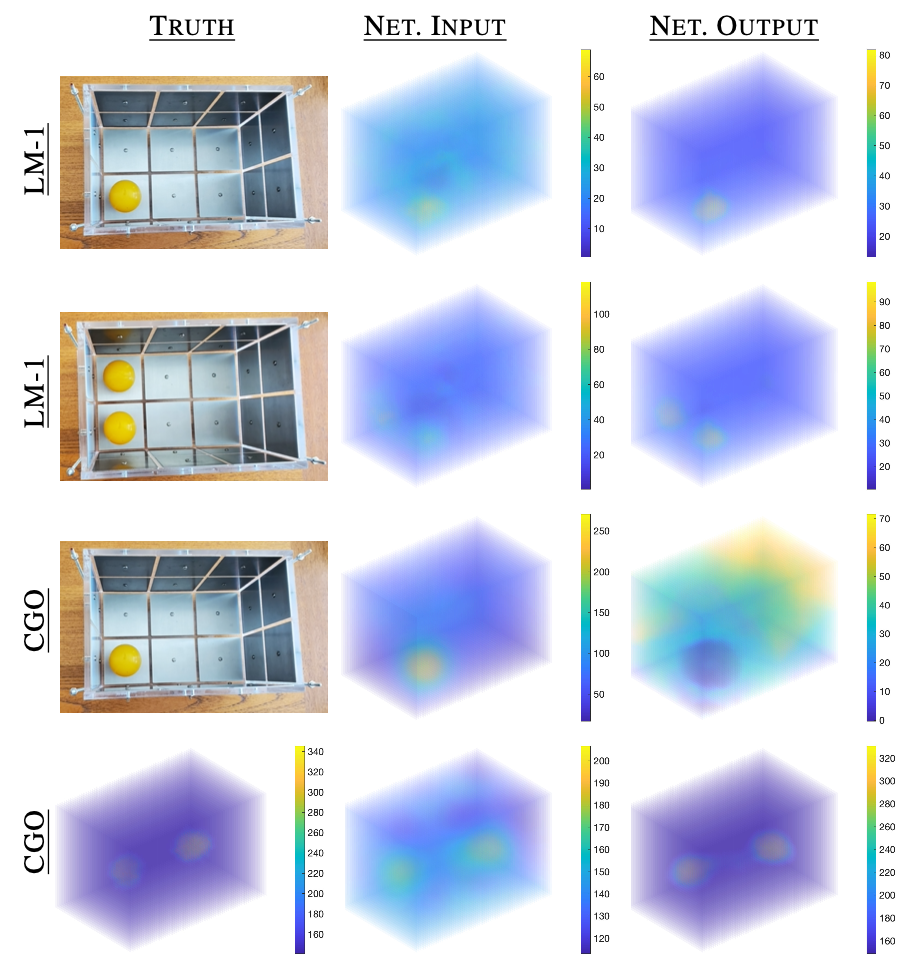}
    \caption{Post-processed reconstructions from experimental ACT5 data from LM iteration 1 and $\texp$ are shown in rows 1-3.  Row~4 shows the result of post-processing a CGO reconstruction in line with the contrast expected by the U1 network using simulated, noisy voltage data.}
	\label{fig:3Dresults}
\end{figure}

The data is defined over a graph, not a pixel grid, as such the convolutions and pooling in the graph U-net are not dependent on the 2D geometry on which the data was trained.  This flexibility allows us to input data from different dimensions.  We take the networks trained on the 2D EIT data in a chest-shaped 16 electrode tank with adjacent current pattern injection, and moderate conductivity contrasts and test how well they generalize to 3D EIT reconstructions from the experimental ACT5 tank data with 32 large electrodes, with different current patterns, and high contrast targets (12x contrast). Here, the inputs to the network are coming from the first iteration of a Levenberg-Marquardt (LM) algorithm, with update term
\begin{equation}\label{eq_update_LM}
\delta\sigma_k^{\scriptscriptstyle (LM)} = - \left( J_k^T J_k + \lambda_{\scriptscriptstyle LM} I \right)^{-1} J_k^T \left( U_k - V \right),
\end{equation}
using $\lambda_{\scriptscriptstyle LM}=1e-6$.  The computational mesh for the 3D box tank shown in Fig.~\ref{fig:ACT5exper} had 85,699 elements and 18,569 nodes.  Solutions were computed on the elements, using linear basis functions, and thus the associated graph had 85,699 nodes.  
Next, the LM iteration 1 reconstructions were scaled by 1/5 to bring the background conductivity value into the window expected by the network.  The 3D reconstructions were then processed through GNN-TV1 and scaled back by 5 yielding the images in Fig.~\ref{fig:3Dresults} (rows 1-2) resulting in significantly sharper images.  The 3D reconstructions are visualized here by stacking transparent slices in the $xy$-plane to render a transparent 3D image.  The images do not achieve the contrast of the true targets, but as the network was not expecting data at 12x contrast, this is not unexpected and in fact is in line with the regularized TV results in Fig.~3 of \cite{Hamilton2022}.  Computational cost of the single iterate was under 10 minutes, not optimized and the post-processing negligible.

Next, we further test the limits by post-processing reconstructions from a direct Complex Geometrical Optics (CGO) based method, the $\texp$ approximation \cite{Hamilton2022,Hamilton2021}, based on the full nonlinear direct solution method \cite{Nachman1988}.  The method is fast (a few seconds) and is based on scattering transforms for the associated Schr\"odinger problem, essentially nonlinear Fourier transforms tailor made for the EIT problem.  For simplicity, the $\texp$ conductivity reconstructions in Fig.~3 of \cite{Hamilton2022}, which can be computed on any type of mesh, were interpolated to the same 85,699 element FEM mesh, scaled again by 1/5, and then used as input to the network GNN-TV1.  As the $\texp$ reconstructions were able to achieve the correct experimental 12x contrast before post-processing, the contrast of the targets input to the network were very far out (6x higher) from what was expected.  As such, the network struggled with what to do with this contrast as can be seen in Fig.~\ref{fig:3Dresults} (row 3).  However, row~4 suggests that the contrast really was the problem, as $\texp$ reconstructions from simulated noisy voltage data corresponding to the contrast expected by the network is post-processed extremely well.  Note how different the artifacts are in the input $\texp$ images when compared to the 2D TV input images.  Nevertheless, the graph U-net is able to sharpen this image remarkably well and adjust the contrast.   

This underlines the flexibility of the graph structure, where training a graph U-net on 2D data and using on 3D data may be particularly advantageous for computationally demanding problems with dense 3D meshes. 

\subsection{Modifying the Network Architecture}\label{sec:netMods}
The growing number of applications utilizing graph data has motivated the increased interest in GNNs over the past several years \cite{Wu2021}. Consequently, a variety of network architectures have been proposed that leverage different geometric aspects of the available graph data. 
While the presented study here considers the suitability of the graph U-net for the post-processing task, we also note, that a possible shortcoming of the graph convolutional layer \cite{KipfWelling2017} is a reduced fitting capacity compared to a standard convolutional layer. Specifically, the graph convolutional layer aggregates information according to a function of the adjacency matrix in contrast to a {\em learned} linear combination of neighboring pixels in the standard convolution. While we presented a convolutional U-Net architecture here, other architectures and layers can be considered. For instance, further improvements to the presented application here could be achieved by using more expressive layers, such as graph attention networks \cite{Velickovic2017} and variants \cite{Wang2020}, or layers exploiting specific geometric relations \cite{Suk2023}.


\subsection{Further Applications}
Aside from the post-processing tasks in the EIT reconstruction problem, other imaging (or non-imaging) inverse problems that utilize irregular and sample-specific domains may benefit from using GNNs. One example is in omnidirectional or $360^{\circ}$ imaging tasks where placing the image on a rectangular pixel grid causes distortion \cite{Su2019}. The use of GNNs instead of CNNs would remove the need for projections and allow the images to remain in their natural spherical domains.  In particular, if only the improved solution at the boundary of an $N$-dimensional object is desired, the solution could be post-processed on a graph consisting only of boundary mesh elements removing the need to project to lower dimensional pixel grids and using determine wrapping conditions in the projected domain.  Additionally, ResNets in model-based learning (e.g. \cite{Herzberg2021})  may be replaced by graph U-nets counterparts, offering larger receptive fields which may be needed to negate nonlocal artifacts in the image or data domain \cite{hauptmann2018approximate,hauptmann2023model}.



\subsection{Python Package GN4IP}
A Python package, Graph Networks for Inverse Problems (GN4IP), was developed to more easily implement learned model-based \cite{Herzberg2021} and post-processing reconstruction methods \cite{HerzbergThesis2022}.  In general, it contains methods for loading datasets from \texttt{.mat} files; building a GNN and CNN with simple sequential or U-net architectures; training and saving model parameters; and predicting; among other things. The package utilizes the PyTorch and PyTorch Geometric libraries for their neural network capabilities. Also, the package is capable of calling on EIDORS\footnote{Electrical Impedance Tomography and Diffuse Optical Tomography Reconstruction Software (EIDORS) is available at \url{eidors3d.sourceforge.net}}, a set of open source algorithms for EIT implemented in MATLAB, to solve the EIT forward problem and compute LM and TV updates needed for the model-based methods.  The GN4IP package is currently available on github\footnote{Graph Networks for Inverse Problems (GN4IP) is available at \url{github.com/wherzberg/GN4IP}}. 

\section{Conclusion}\label{sec:conclusions}
A new graph U-net alternative to the convolutional U-net that has been used extensively for imaging tasks was presented here. The presented architecture is distinguished by the proposed \textit{k-means cluster max pool layer} and \textit{clone cluster unpool layer}.  The k-means cluster max pool layer behaves similarly to the max pool layer of the convolutional U-net in that it aggregates features within local windows of the input's data structure. This was different from previously used hierarchical, node-dropping layers. The clone cluster unpool layer works naturally with the cluster assignments of the associated k-means cluster max pool layer to up-sample the input graph by restoring the original graph structure. The main advantage of the presented graph U-net over the CNN U-Net is given by the flexibility provided by the graph framework, allowing application to irregular data defined over FEM meshes and being dimension agnostic.

Using EIT as a case study, the new graph U-net was tested on six different experimental datasets coming from three different EIT machines both in 2D and 3D. Compared to the classic CNN alternative, the proposed network shows comparable performance and the k-means cluster maxpool layers provide similar behaviour.  The advantage of the graph framework comes with the added flexibility to process on irregular data where interpolation between meshes is undesirable or computationally expensive. A significant advantage of the graph, as presented, is the ability to train on lower dimensional data (e.g. 2D) and application to higher dimensional data (e.g. 3D).  
This is a conceptual difference to the CNN pixel/voxel based setting, where filters are dimension dependent.

Compared to the full iterative TV method used as baseline, the presented post-processing uses only the first few iterates, effectively reducing the inference time. 
On average, each iteration of the TV method took about 1.7 seconds per sample in 2D, while the application of a trained U-net of either type took only 0.01 seconds per sample. Therefore, eliminating a fraction of the required iterations reduces the inference time by about the same fraction, i.e., 20 versus 4 iterates or less. The time savings become even more valuable in 3D applications where the meshes contain more elements and each iteration of the classical method of choice takes considerably longer, often even computationally prohibitive. 

In regards to inverse problems in general, GNNs have been demonstrated to be a fast, flexible, and interpretable option for applying deep learning. Our work here indicates they are a viable, and possibly, superior alternative to other network types. Continued research on novel GNN layers, architectures, and applications is encouraging for the future of GNNs for inverse problems.

\section*{Acknowledgement} 

We gratefully acknowledge the support of NVIDIA Corporation with the donation of the Titan Xp GPUs used for this research as well as the Raj high performance cluster at Marquette University funded by the National Science foundation award CNS-1828649.  
We also thank the EIT groups at RPI and UEF for sharing the respective experimental data sets. 
\bibliographystyle{IEEEtran}
\small
\bibliography{IEEEabrv,bib_clean, bib}

\end{document}